\documentclass{article}

\newif\ifusenixsty
\usenixstytrue
\newif\ifcameraready
\camerareadyfalse

\pdfoutput=1

\usepackage{lmodern}
\usepackage{amssymb,amsmath,amsfonts}
\usepackage[T1]{fontenc}
\usepackage[utf8]{inputenc}

\IfFileExists{upquote.sty}{\usepackage{upquote}}{}

\IfFileExists{microtype.sty}{%
 \usepackage{microtype}
 \UseMicrotypeSet[protrusion]{basicmath} 
}{}

\usepackage{hyperref}
\usepackage{natbib} 
\usepackage{graphicx,grffile}
\usepackage[accepted]{sysml2019}

\usepackage{endnotes,epsfig}
\usepackage{multirow}
\usepackage{xspace}
\usepackage{tabularx}
\usepackage{ragged2e}
\usepackage{booktabs}
\usepackage{paralist}
\usepackage[american]{babel}
\usepackage[shortlabels]{enumitem}
\usepackage{courier,color,wrapfig}
\usepackage{xspace}
\usepackage{balance}
\usepackage{enumitem}
\usepackage{epstopdf}
\usepackage{multirow}
\usepackage{booktabs}
\usepackage{url}
\usepackage{amssymb}
\usepackage{pifont}
\usepackage{subfigure}
\usepackage{tikz}
\usepackage{comment}
\usepackage{mathtools}
\usepackage{ulem}
\usepackage{xkeyval}
\usepackage{ifthen}
\usepackage{etoolbox}
\usepackage[margin=5pt,font=small,labelfont={bf,rm}]{caption}
\usepackage[compact]{titlesec}
\usepackage{bbding}
\usepackage{enumitem}

\usepackage{array}
\newcolumntype{L}[1]{>{\raggedright\let\newline\\\arraybackslash\hspace{0pt}}m{#1}}
\newcolumntype{C}[1]{>{\centering\let\newline\\\arraybackslash\hspace{0pt}}m{#1}}
\newcolumntype{R}[1]{>{\raggedleft\let\newline\\\arraybackslash\hspace{0pt}}m{#1}}
\usepackage{amsmath}

\PassOptionsToPackage{usenames,dvipsnames}{color} 
\hypersetup{unicode=true,
            colorlinks=true,
            linkcolor=blue,
            citecolor=blue,
            anchorcolor=blue,
            urlcolor=blue,
            breaklinks=true}
\makeatletter
\def\maxwidth{\ifdim\Gin@nat@width>\linewidth\linewidth\else\Gin@nat@width\fi}
\def\maxheight{\ifdim\Gin@nat@height>\textheight\textheight\else\Gin@nat@height\fi}
\makeatother
\setkeys{Gin}{width=\maxwidth,height=\maxheight,keepaspectratio}
\makeatletter
\let\origsection\section
\let\origsubsection\subsection

\renewcommand\section{\@ifstar{\starsection}{\nostarsection}}
\renewcommand\subsection{\@ifstar{\starsubsection}{\nostarsubsection}}

\newcommand\sectionprelude{\vspace{0ex}}
\newcommand\sectionpostlude{\vspace{0ex}}
\newcommand\subsectionprelude{\vspace{0ex}}
\newcommand\subsectionpostlude{\vspace{0ex}}

\newcommand\nostarsection[1]{\sectionprelude\origsection{#1}\sectionpostlude}
\newcommand\starsection[1]{\sectionprelude\origsection*{#1}\sectionpostlude}

\newcommand\nostarsubsection[1]{\subsectionprelude\origsubsection{#1}\subsectionpostlude}
\newcommand\starsubsection[1]{\subsectionprelude\origsubsection*{#1}\subsectionpostlude}

\makeatother

\newcommand\paraspace{\vspace*{0ex}}
\providecommand\parab[1]{\paraspace\noindent\textbf{#1}.}
\providecommand\parae[1]{\textbf{\textit{#1}}.}

\apptocmd\normalsize{%
\abovedisplayskip=5pt
\abovedisplayshortskip=5pt
\belowdisplayskip=5pt
\belowdisplayshortskip=5pt
}{}{}

\hypersetup{final}

\newcommand{\sysname}{Satyam\xspace}
\newcommand{\satyam}{Satyam\xspace}

\newcommand{\etc}{\textit{etc.}\xspace}
\newcommand{\ie}{\textit{i.e.,}\xspace}
\newcommand{\eg}{\textit{e.g.,}\xspace}

\newcommand{\secref}[1]{\S\ref{#1}}
\newcommand{\figref}[1]{Figure~\ref{#1}}

\newcommand{\squishlist}{
 \begin{list}{$\bullet$}
 		{ \setlength{\itemsep}{0pt}
 			\setlength{\parsep}{3pt}
 			\setlength{\topsep}{3pt}
 			\setlength{\partopsep}{0pt}
 			\setlength{\leftmargin}{1.5em}
 			\setlength{\labelwidth}{1em}
 			\setlength{\labelsep}{0.5em} } }

\newcommand{\squishend}{
  \end{list}  }

\ifcameraready

\fi

\sysmltitlerunning{Satyam}

\begin{document}

\twocolumn[
\sysmltitle{Satyam: Democratizing Groundtruth for Machine Vision}

\vspace{-3mm}
\protect\parbox{\textwidth}{\protect\centering Hang Qiu\textsuperscript{$\star$}, Krishna Chintalapudi\textsuperscript{$\dag$}, Ramesh Govindan\textsuperscript{$\star$}\\ $^\star$University of Southern California \hspace{1cm} $^\dag$Microsoft Research}

\vspace{1mm}

\begin{abstract}
The democratization of machine learning (ML) has led to ML-based machine vision systems for autonomous driving, traffic monitoring, and video surveillance. However, true democratization cannot be achieved without greatly simplifying the process of collecting groundtruth for training and testing these systems. This groundtruth collection is necessary to ensure good performance under varying conditions.
In this paper, we present the design and evaluation of \textbf{\satyam}, a first-of-its-kind system that enables a layperson to launch groundtruth collection tasks for machine vision with minimal effort. \satyam leverages a crowdtasking platform, Amazon Mechanical Turk, and automates several challenging aspects of  groundtruth collection: creating and launching of custom web-UI tasks for obtaining the desired groundtruth, controlling result quality in the face of spammers and untrained workers, adapting prices to match task complexity, filtering spammers and workers with poor performance, and processing worker payments. We validate  \satyam using several popular benchmark vision data sets, and demonstrate that groundtruth obtained by \satyam is comparable to that obtained from trained experts and provides matching ML performance when used for training. \looseness=-1
\end{abstract}
]



\section{Introduction}
\label{sec:intro}

The accuracy of deep neural network based machine vision systems depends on the groundtruth data used to train them. Practically deployed systems rely heavily on being trained and tested on groundtruth data from images/videos obtained from actual deployments. Often, practitioners start with a model trained on public data sets and then fine-tune the model by re-training the last few layers using groundtruth data on images/videos from the actual deployment~\cite{retrain1,retrain2} in order to improve accuracy in the field. \looseness=-1

Obtaining groundtruth data, however, can present a significant barrier, as annotating images/videos often requires significant human labor and expertise. Today, practitioners use three different approaches to groundtruth collection. Large companies employ their own trained workforce to annotate groundtruth. Third parties such as \cite{spare5_website} and \cite{crowdflower} recruit, train and make available a trained workforce for annotation. Finally, \textit{crowdtasking platforms} such as Amazon Mechanical Turk (AMT~\cite{AMT}) provide rapid access to a pool of untrained \textit{workers} that can be leveraged to generate groundtruth annotations. 

While the first two approaches have the advantage of generating high quality groundtruth by using a trained workforce, they incur significant cost in recruitment and training, and are therefore often limited to well-funded companies. Consequently, employing a crowdtasking platform like AMT is often a preferred alternative for a large number of ML practitioners. Using AMT for obtaining groundtruth, however, presents several challenges that deter its widespread use. First, requesters may not always have the expertise needed to create user-friendly web user-interfaces to present to workers for annotation tasks. Second, worker quality varies widely in AMT, and results can be corrupted by spammers and bots, so requesters must curate results manually to obtain good groundtruth. Third, machine vision often requires groundtruth for hundreds or thousands of images and videos, and generate AMT Human Intelligence Tasks (HITs) manually is intractable, as is determining which workers need to be paid, or which workers to recruit.





\parab{Goal, Approach and Contributions}
In this paper, we ask: \textit{Is it possible to design a groundtruth collection system that is accessible to non-experts while also being both cost-effective and accurate?} To this end, we discuss the design, implementation, and evaluation of \satyam\footnote{The Satyam portal~\cite{satyamportal_website} is functional, but has not yet been released for public use.} (\secref{sec:overview}) which allows non-expert users to launch large groundtruth collection tasks in under a minute. 

\satyam users first place images/video clips at a cloud storage location. They then specify groundtruth collection declaratively using a web-portal. After a few hours to a few days (depending on the size and nature of the job), \satyam generates the groundtruth in a consumable format and notifies the user. Behind the scenes, in order to avoid the challenges of recruiting and managing trained annotators, \satyam leverages AMT workers, but automates generation of customized web-UIs, quality control and HIT management.

\begin{figure}[t]
  \centering
 \includegraphics[width=\columnwidth]{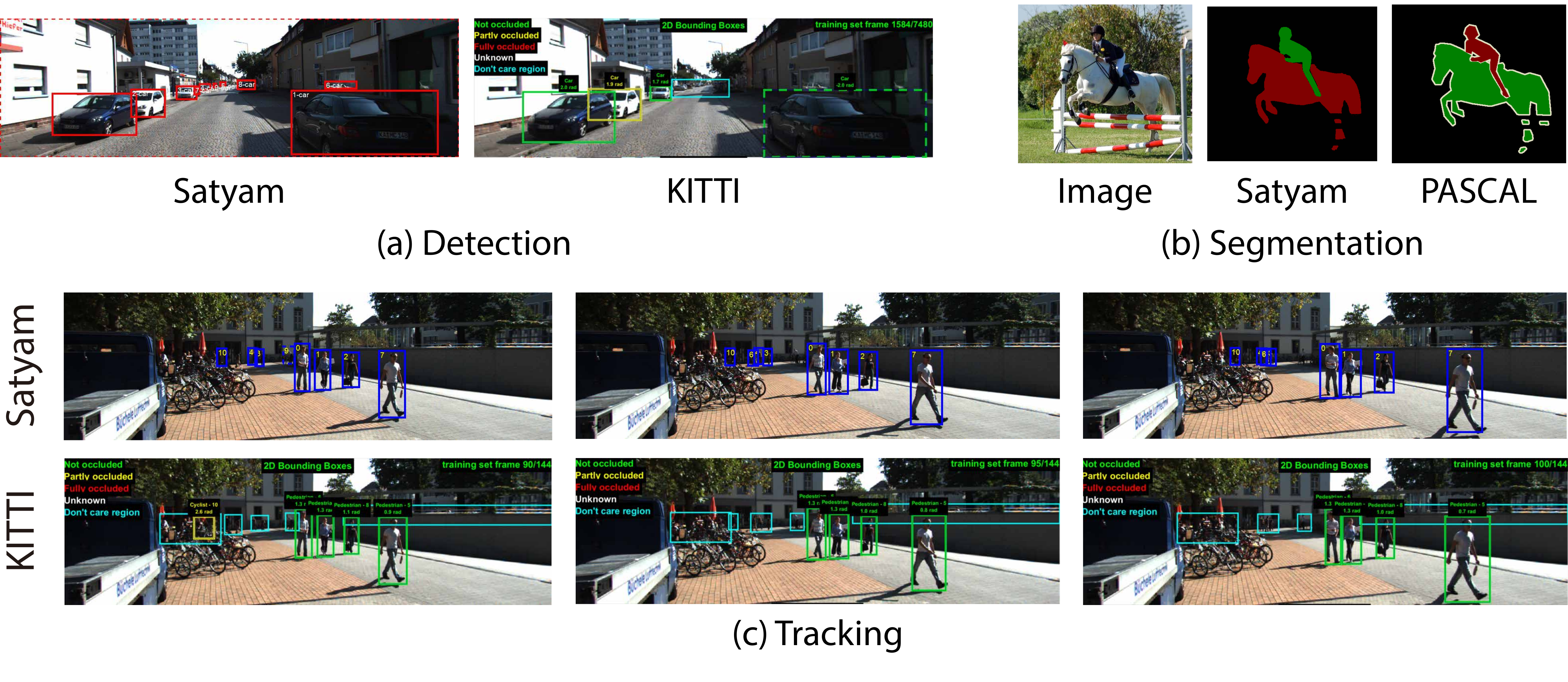}
  \vspace*{-0.15in}
  \caption{Examples of \satyam Results on Detection, Segmentation, and Tracking.}
  \label{fig:satyamresults}
\end{figure}

\parae{High-level Specification} 
A key challenge in using AMT arises because the HIT is too low-level of an abstraction for large-scale groundtruth collection. \satyam elevates the abstraction for groundtruth collection by observing that machine vision tasks naturally fall into a small number of categories (\secref{sec:templates}), \eg classification (labeling objects in an image or a video), detection (identifying objects by drawing bounding boxes), segmentation (marking pixels corresponding to areas of interest) and a few others described in \secref{sec:templates}. \satyam allows users to specify their groundtruth requirements by providing customizable specification templates for each of these tasks. \looseness=-1


\parae{Automated Quality Control} 
\satyam automates quality control in the face of an untrained workforce and eliminates the need for manual curation (\secref{sec:fusion}). It requests annotations from multiple workers for each  image/video clip. Based on the assumption that different workers make independent errors in the groundtruth annotation, \satyam employs novel {\it groundtruth-fusion} techniques that identify and piece together the ``correct parts'' of the annotations from each worker, while rejecting the incorrect ones and requesting additional annotations until the fused result is satisfactory. 

\parae{Automated HIT Management - Pricing, Creation, Payment and Worker Filtering} 
\satyam automates posting HITs in AMT for each image/video in the specified storage location until the groundtruth for that image/video has been obtained. Instrumentation in \satyam's annotation web-UIs allow it to measure the time taken for each HIT. \satyam uses this information to adaptively adjust the price to be paid for various hits and ensures that it matches the requester's user-defined hourly wage rate.   \satyam determines whether or not a worker deserves payment by comparing their work against the final generated groundtruth and disburses payments to deserving workers. When recruiting workers, it uses past performance to filter out under-performing workers.

\parab{Implementation and Deployment} \satyam's implementation is architected using a collection of cloud functions that can be auto-scaled, that support asynchronous result handling with humans in the loop, and that can be evolved and extended easily. Using an implementation of \satyam on Azure, we evaluate (\secref{sec:eval}) various aspects of \sysname. We find that \satyam's groundtruth almost perfectly matches groundtruth from well known publicly available image/video data sets such as KITTI \cite{KITTI} which were annotated by experts or by using sophisticated equipment. Further, ML models re-trained using \satyam groundtruth perform identically with the same models re-trained with these benchmark datasets. We have used \satyam for over a year launching over 162,000 HITs on AMT to over 12,000 unique workers.

\parab{Examples of groundtruth generated by \satyam}
\figref{fig:satyamresults} show examples of the groundtruth generated by \satyam for detection, segmentation, and tracking, and how these compare with groundtruth from benchmark datasets. More examples are available in Figures~\ref{fig:DetResult},~\ref{fig:SegResult} and~\ref{fig:TrackingResult} and at~\cite{Tracking_samplesdropboxlink,Localization_samplesdropboxlink}.

\section{\sysname Overview}
\label{sec:overview}

\sysname is designed to minimize friction for non-expert users when collecting groundtruth for ML-based machine vision systems. It uses AMT but eliminates the need for users to develop complex Web-UIs or manually intervene in quality control or HIT management.





\subsection{Design Goals}
\label{sec:overview:design_goals}
We now briefly describe the key design goals that shaped the architectural design of \satyam.

\parab{Ease of use} \satyam's primary design goal is \textbf{\textit{ease of use}}. \satyam users should only be required to provide a high-level declarative specification of ground-truth collection: what kind of ground-truth is needed (\eg class labels, bounding boxes), for which set of images/videos, how much the user is willing to pay, \etc \satyam should not require user intervention in designing UIs, assessing result quality, or ensuring the appropriate HIT price, \etc, but must perform these automatically. 

\parab{Scalability} \satyam will be used by multiple concurrent users, each running several different ground-truth collection activities. Each activity in turn might spawn several tens of thousands of requests to workers, and each request might involve generating web user interfaces, assessing results, and spawning additional requests, all of which might involve significant compute and storage.

\parab{Asynchronous Operation} Because \satyam relies on humans in the loop, its design needs to be able to \textbf{\textit{tolerate large and unpredictable delays}}. Workers may complete some HITs within a few seconds of launch, or may take days to process HITs. 

\parab{Evolvability} In designing algorithms for automating ground truth collection, \satyam needs to take several design dimensions into account: the requirements of the user, the constraints of the underlying AMT platform, variability in worker capabilities, and the complexity of assessing visual annotation quality. These algorithms are complex, and will evolve over time, and \satyam's design must accommodate this evolution.


\parab{Extensibility}
ML for machine vision is rapidly evolving and future users will need new kinds of groundtruth data which \satyam must be able to accommodate.

\looseness=-1

\begin{figure}[t]
\centering
\begin{minipage}{0.9\columnwidth}
   \centering\includegraphics[width=0.7\textwidth]{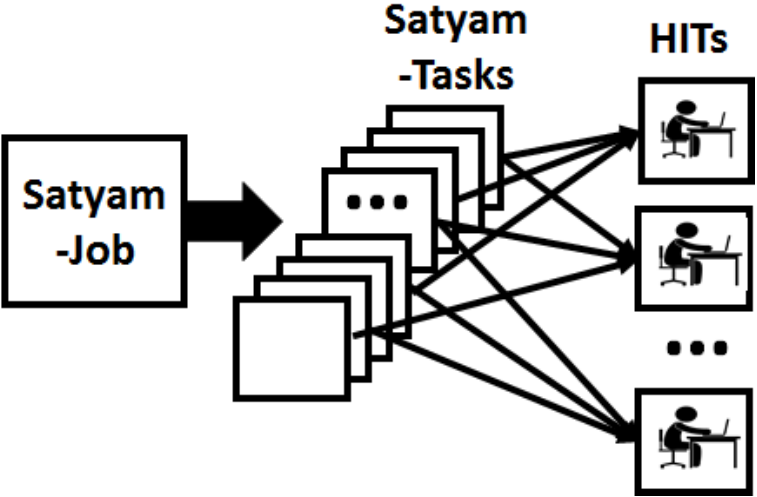}
   \caption{\sysname's jobs, tasks and HITs}
   \label{fig:JobsTasksHits}
\end{minipage}
\end{figure}

\begin{figure*}[t]
  \centering
  \begin{minipage}{0.75\columnwidth}
    \centering
    \includegraphics[width=0.95\textwidth]{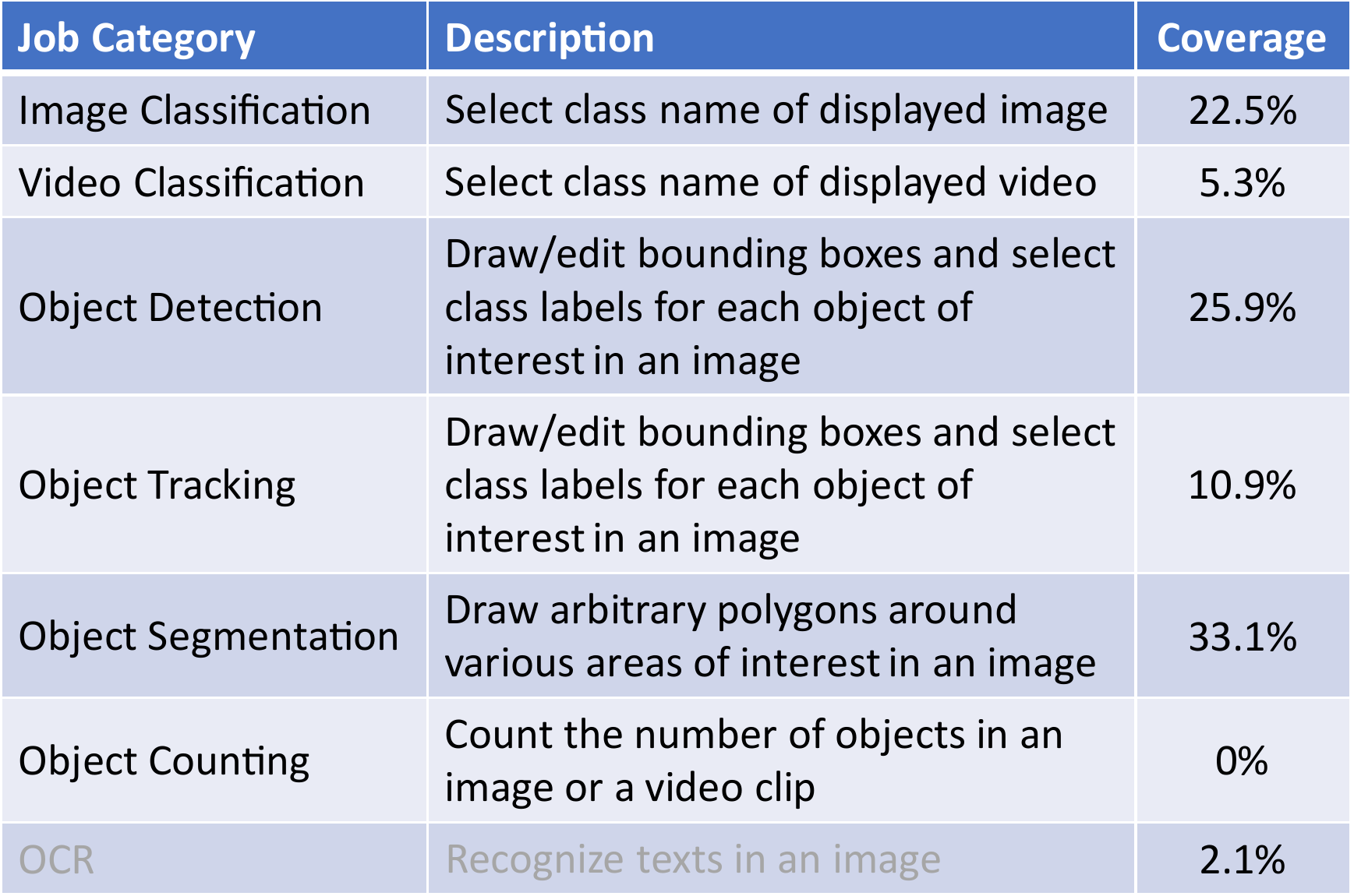}
    \caption{\satyam Job Categories}
    \label{fig:templates}
  \end{minipage}
  \begin{minipage}{1.25\columnwidth}
  \centering\includegraphics[width=0.90\textwidth]{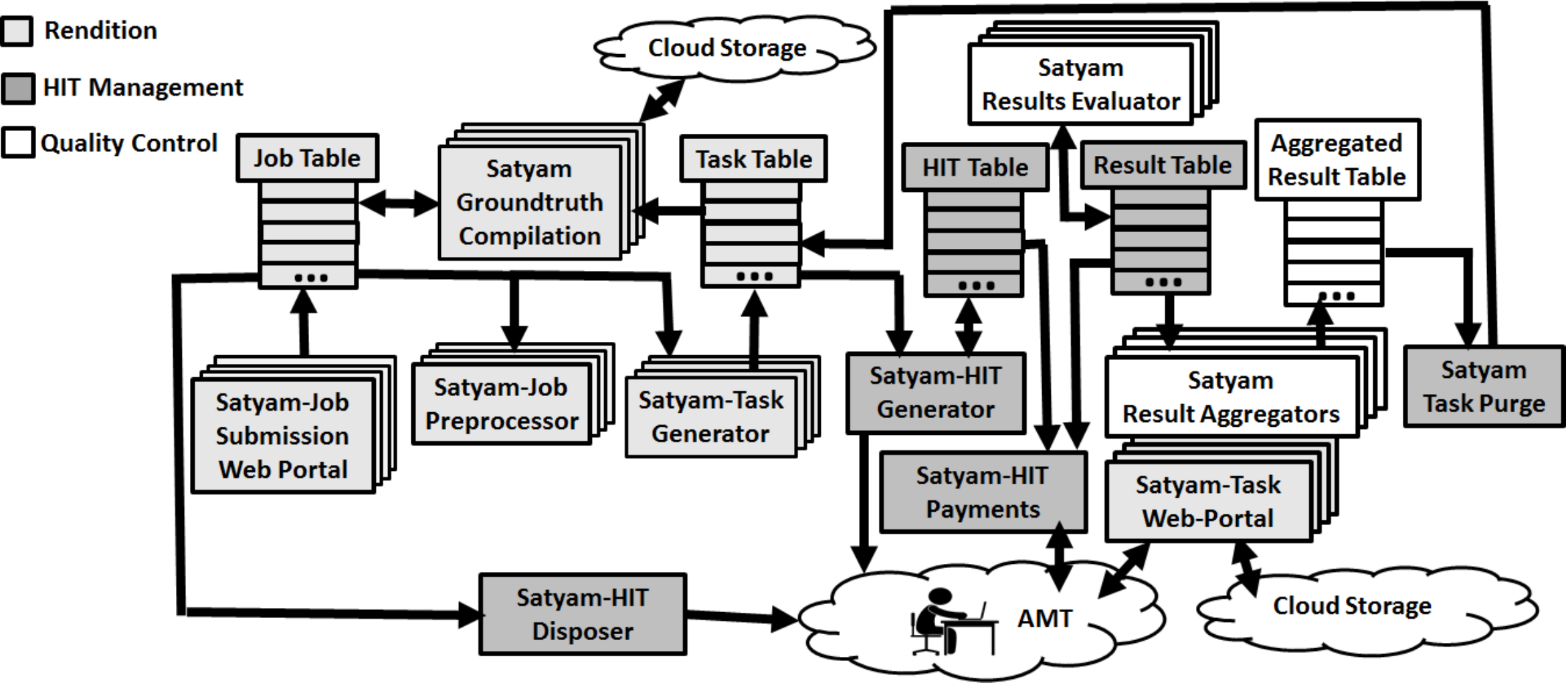}
  \caption{Overview of the \sysname's components}
  \label{fig:Satyam-Components}
  \end{minipage}
  \vspace*{-2mm}
\end{figure*} 


\subsection{\satyam Abstractions}
\label{sec:overview:abstractions}
\satyam achieves ease of use and asynchronous operation by introducing different abstractions to represent logical units of work in groundtruth annotation (depicted in Figure~\ref{fig:JobsTasksHits}).

\parab{\satyam-job} Users specify their groundtruth collection requirements at a high-level as a \textbf{\textit{\satyam-job}}, which has several parameters: the set of images/video clips, the kind of groundtruth desired (\eg bounding rectangles for cars), payment rate (\$/image or  \$/hour), the AMT requester account information to manage HITs on the user's behalf, \etc At any instant, \satyam might be running multiple \satyam-jobs.

\parab{\satyam-tasks} \satyam renders jobs to \textbf{\textit{\satyam-tasks}}, which represent the smallest unit of groundtruth work sent to a worker. For example, a \sysname-task might consist of a single image in which a worker is asked to annotate all the bounding rectangles and their classes (\secref{sec:intro}), or a short video clip in which a worker is asked to track one or more objects. A single \sysname-job might spawn hundreds to several tens of thousands of \sysname-tasks.  

\parab{HIT} A \textbf{\textit{HIT}} is an AMT abstraction for the smallest unit of work for which a worker is paid. \satyam decouples HITs from \satyam-tasks, for two reasons. First, \satyam may batch multiple \satyam-tasks in a HIT. For example, it might show a worker 20 different images (each a different \sysname-task) and ask her to classify the images as a part of a single HIT. Batching increases the price per HIT thereby incentivizing workers more, and also increases worker throughput. Second, it allows a single \satyam-task to be associated with multiple HITs, one per worker: this permits \satyam to obtain groundtruth for the same image from multiple workers to ensure high quality results. \looseness=-1

 
\subsection{\satyam Architecture}
\label{sec:overview:architecture}
\sysname is architected as a collection of components (\figref{fig:Satyam-Components}) each implemented as a \textit{cloud function} (\eg an Azure function or an Amazon lambda) communicating through persistent storage. This design achieves several of \satyam's goals. Each component can be scaled and evolved independently. Components can be triggered by users requesting new jobs or workers completing HITs and can thereby accommodate asynchronous operation. Finally, only some components need to be modified in order to extend \satyam to new types of ground truth collection.

\satyam's components can be grouped into three high-level functional units, as shown in \figref{fig:Satyam-Components}: {\it Job Rendition}, {\it Quality Control} and {\it HIT Management}. We describe these components, and their functional units, in the subsequent sections.

 \parab{Job Rendition} 
 This functional unit raises the level of abstraction groundtruth collection (Section~\ref{sec:templates}). It is responsible for translating the user's high level groundtruth collection requirements to AMT HITs and then compiling the AMT worker results into a presentable format for users. Users primarily interact with the {\it Job-Submissions Portal}~\cite{satyamportal_website} where they submit their groundtruth collection requirements. Submitted jobs are written to the {\it Job-Table}. Based on the job descriptions in the Job-Table, the {\it Pre-processor} may perform data manipulations such as splitting videos into smaller chunks. The {\it Task Generator}  decomposes the \sysname-job into \sysname-tasks, one for each image/video chunk. The {\it Task Portal} is a web application that dynamically renders \sysname-tasks into web pages (based on their specifications) displayed to AMT workers. Finally, {\it Groundtruth Compilation} assembles the final results from the workers for the entire job and provides them to \sysname users as a JSON format file.

 \parab{Quality Control}
 AMT workers are typically untrained in groundtruth collection tasks and \sysname has little or no direct control over them. Further, some of the workers might even be bots intending to commit fraud~\cite{mturk-spam}. The quality control components are responsible for ensuring that the groundtruth generated by \sysname is of high quality. In order to achieve this, \sysname sends the same task to multiple non-colluding workers and combines their results. The {\it Result Aggregator} identifies and fuses the ``accurate parts'' of workers' results while rejecting the ``inaccurate parts'' using {\it groundtruth fusion algorithms} described in \secref{sec:fusion}. For certain tasks, the Aggregator might determine that it requires more results to arrive at a conclusive high quality result. In that case, it presents the task to more workers until a high quality groundtruth is produced. The {\it Results Evaluator} compares fused results with the individual worker's results to determine whether the worker performed acceptably or not and indicates this in the {\it Result-Table}. \looseness=-1

\parab{HIT-Management} These components directly interact with the Amazon AMT platform and manage HITs through their life-cycle (\secref{sec:payments}). The {\it HIT Generator} reads the task table and launches HITs in AMT and always ensures that there are no unfinished tasks with no HITs. It is also responsible for {\it adaptive pricing} -- adaptively adjusting the HIT price by measuring the median time to task completion, and {\it worker filtering} -- ensuring that under-performing workers are not recruited again.  The {\it HIT Payments } component reads the results table and pays workers who have completed a task acceptably, while rejecting payments for those who have not. The {\it Task Purge} component removes tasks from the task table that have already been aggregated, so that they are not launched as HITs again and the {\it HIT Disposer} removes any pending HITs for a completed job.

\section{Job Rendition}
\label{sec:templates}

To achieve ease of use, \satyam needs to provide users with an expressive high-level specification framework for ground truth collection. \satyam leverages the observation that, in the past few years, the machine vision community has organized its efforts around a few well-defined categories of  \textit{vision tasks}: classification, detection or localization, tracking, segmentation, and so forth. \satyam's job specification is based on the observation that different ground truth collection within the same vision task category (\eg classification of vehicles vs. classification of animals) share significant commonality, while ground-truth collection for different vision task categories (\eg classifying vehicles vs. tracking vehicles) are qualitatively different.


\parab{Job Categories}
\satyam defines a small number of \textit{job categories} where each category has similar ground-truth collection requirements. Users can customize groundtruth collection by parameterizing a job category \textit{template}. For example, to collect class label groundtruth for vehicles (\eg car, truck, \etc), a user would select an image classification job template and specify the various vehicle class labels.

Templatizing job categories also enables \satyam to automate all steps of ground-truth collection. The web UIs presented to AMT workers for different ground truth collection jobs in the same category (\eg classification) are similar, so \satyam can automatically generate these from \textit{Web-UI templates}. Moreover, quality control algorithms for ground truth collection in the same category are similar (modulo simple parametrization), so \satyam can also automate these. \looseness=-1


To determine which job categories to support, we examined the top 400 publicly available groundtruth datasets used by machine vision researchers~\cite{yacvid} and categorized them with respect to the Web-UI requirements for obtaining the groundtruth (\figref{fig:templates}). The \textit{coverage} column indicates the fraction of datasets falling into each category. \satyam currently supports the first six categories in \figref{fig:templates}, which together account for the groundtruth requirements of more than 98.1\% of popular datasets in machine vision. We now briefly describe a few of the most used currently available templates in \satyam.

\begin{figure}
  \centering
  \includegraphics[width=\columnwidth]{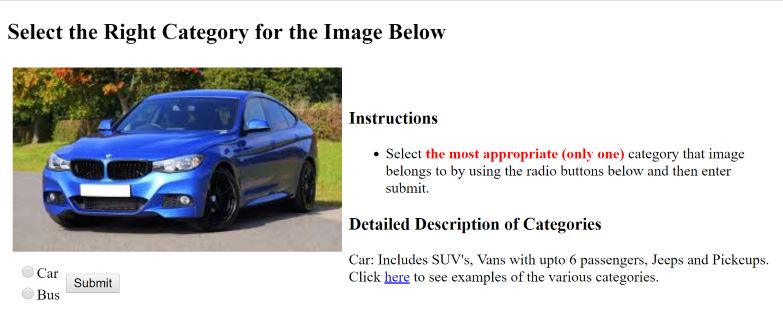}
  \caption{Image Classification Task Page}
  \label{fig:imClassfication_taskpage}
\end{figure} 

\begin{figure}
  \centering
  \includegraphics[width=\columnwidth]{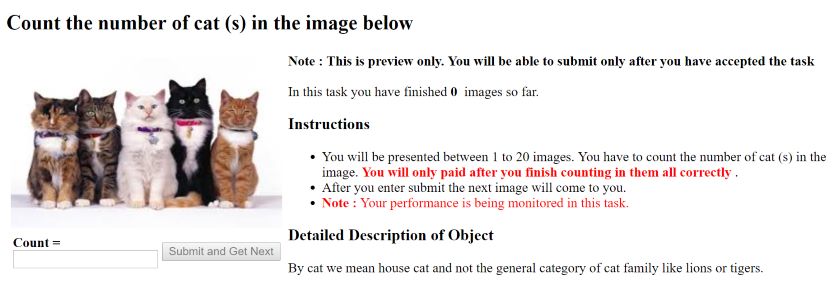}
  \caption{Object Counting Task Page}
  \label{fig:counting_taskpage}
\end{figure}

\parae{Image and Video Classification}
The desired groundtruth in this category is the label (or labels), from among a list of provided class labels, that most appropriately describes the image/video. 
Class labels can describe objects in images such as cars or pedestrians and actions in video clips such as walking, running, and dancing.  
\satyam users customize (\figref{fig:imClassfication_taskpage}) the corresponding job templates by providing the list of class labels and a link or description for them. To the workers, the web-UI displays the image/video clip with the appropriate instructions and a radio button list of class labels.

\begin{figure*}
\centering
\begin{minipage}{0.66\columnwidth}
  \centering
  \includegraphics[width=\columnwidth]{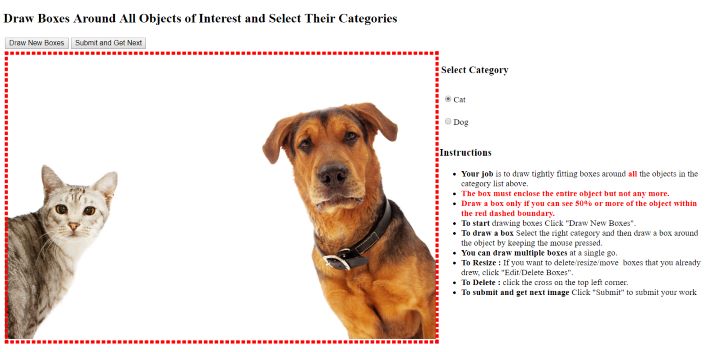}
  \caption{Object Detection and Localization Task Page}
  \label{fig:imDetection_taskpage}
\end{minipage}
\begin{minipage}{0.66\columnwidth}
  \centering
  \includegraphics[width=\columnwidth]{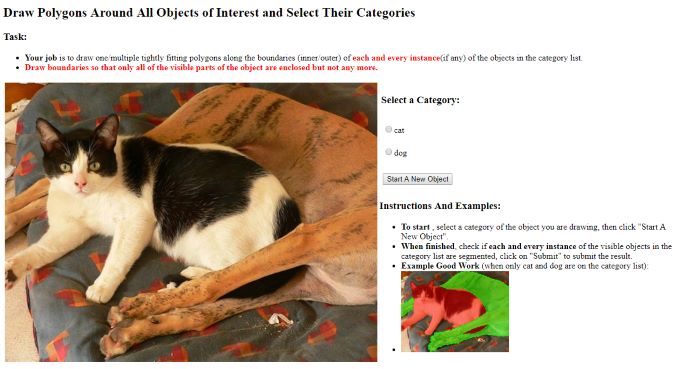}
  \caption{Object Segmentation Task Page}
  \label{fig:imSegmentation_taskpage}
\end{minipage}
\begin{minipage}{0.66\columnwidth}
  \centering
  \includegraphics[width=\columnwidth]{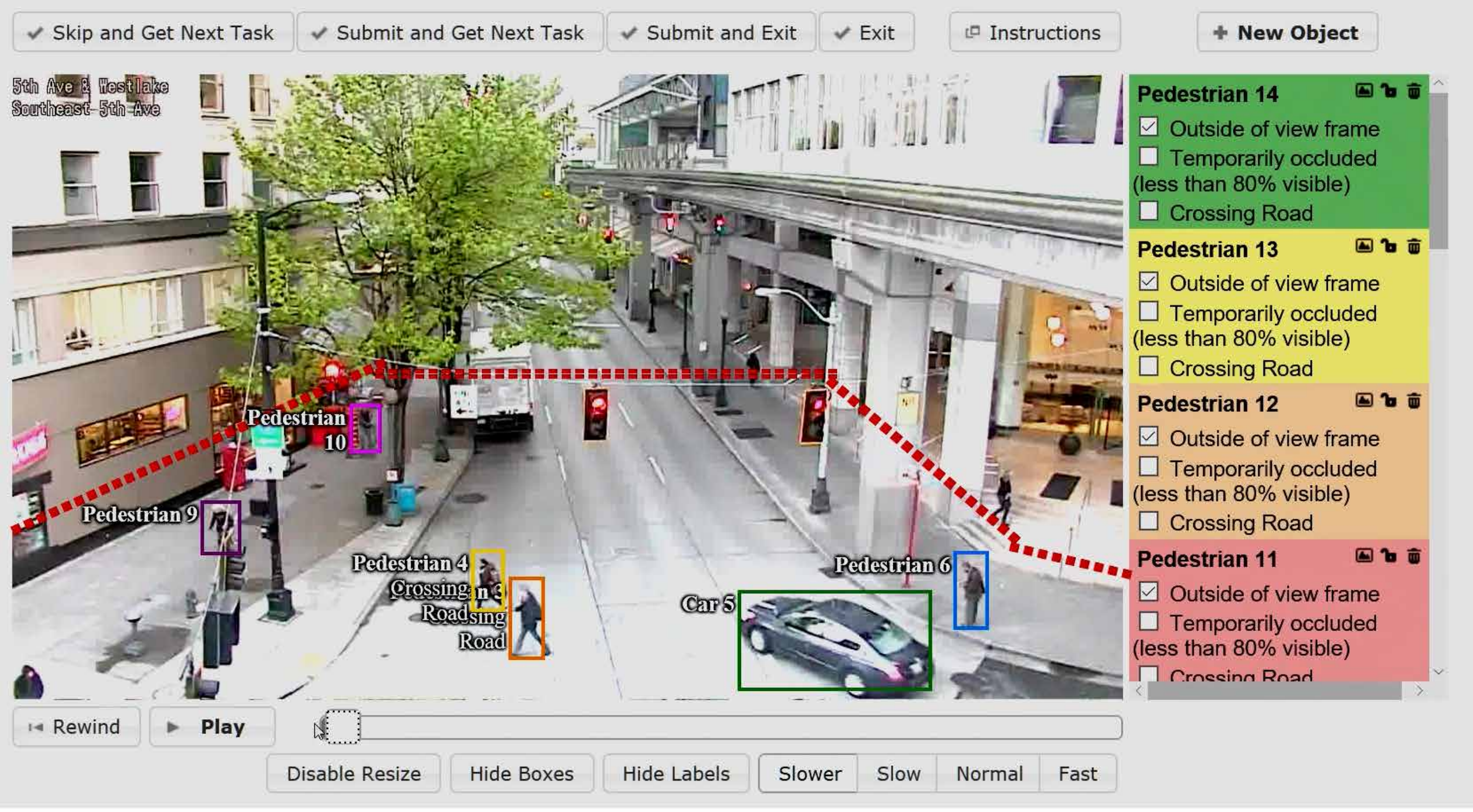}
  \caption{Multi-Object Tracking Task Page}
  \label{fig:tracking_taskpage}
\end{minipage}
\end{figure*}

\parae{Object Counting in Images and Videos}
 The desired groundtruth for this job category is a count of objects of a certain class, or of events in an image or video (\eg the number of cars in a parking lot or the number of people entering a certain mall or airport).  The user provides a description of the object/event. In the web-UI, the worker is shown an image/video clip (\figref{fig:counting_taskpage}), and the description provided of the object/event of interest, for which the worker is asked to provide a count.

\parae{Object Detection in Images}
 The desired groundtruth in this category is a set of bounding boxes on an image marking parts of interest in the image, along with a class label that most appropriately describes each box.  Users specify (\figref{fig:imDetection_taskpage}) the object classes for which workers should draw bounding boxes describing areas within each image that need to be annotated. Workers see an image with a radio button list of object classes, using which the workers can select one class to draw/edit bounding boxes around all objects of the same class, \eg all pedestrians in a traffic surveillance image, in one shot.

For example, in a traffic surveillance scene, the objects of interest might be all the cars and pedestrians. The groundtruth required for such algorithms for each image, is the set of all bounding boxes enclosing the objects of interest and their respective category names.
 To support these cases we provide a template that generates a web-UI where workers are displayed an image and can draw/edit bounding boxes around objects of interest (using the mouse). A radio button list of the categories helps the workers categorize the object as well. 
\sysname users customize this template by specifying the categories of interest. The users may also specify a set of polygons describing the various areas of interest within the images.

\parae{Object Segmentation in Images} The desired groundtruth in this category is pixel-level annotations of various objects/areas of interest (\eg people, cars, the sky). This template is similar to the object detection template except that it lets workers annotate arbitrary shapes by drawing a set of polygons (\figref{fig:imSegmentation_taskpage}). 

\parae{Object Tracking in Videos}
The desired groundtruth in this category, an extension of object detection to videos, requires bounding boxes for each distinct object/event of interest in successive frames of a video clip. This groundtruth can be used to train object trackers. \satyam users can select (\figref{fig:tracking_taskpage}) the video tracking job category, and specify the object classes that need to be tracked, instructions to workers on how to track them, 
 what frame rate the video should be annotated at, and polygons that delineate areas of interest within frames. Workers are presented (\figref{fig:tracking_taskpage}) with a short video sequence, together with the categories of interest, and can annotate bounding boxes for each object on each frame of the video. For annotation,  we have modified an existing open source video annotation tool \cite{vatic} and integrated it into \satyam.


\parab{Job Rendition Components}
When a user wishes to initiate ground truth collection, she uses the \textit{Job Submission Portal} to select a job category template, and fills in the parameters required for that template. Beyond the category specific parameters described above, users provide a cloud storage location containing the images or videos to be annotated, and indicate the price they are willing to pay. After the user submits the job specification, the Portal generates a globally unique ID (GUID) for the job, and stores the job description in the \textit{Job-Table}. Then, the following components perform job rendition.

\parae{Pre-processor} After a job is submitted via the Job Submission portal, the images/video clips might need to be pre-processed. In our current implementation, \sysname supports preprocessing for video annotations. Specifically, large videos (greater than 3 second duration) are broken into smaller chunks (with a small overlap between successive chunks to facilitate reconstruction or \textit{stitching}, see below) to diminish cognitive load on workers. They are then downsampled based on user's requirements, and converted into a browser-friendly format (\eg MP4).

\parae{Task Generator} 
This component creates a \sysname-task for each image or video chunk. A \sysname-task encapsulates all the necessary information (image/video URI, user customizations, the associated Job-GUID, \etc) required to render a web-page for the image/video clip. The \sysname-task is stored as a JSON string in the \textit{Task-Table}. The Task Table stores additional information regarding the task, such as the number of workers who have attempted it.

\begin{figure}
  \centering
  \includegraphics[width=\columnwidth]{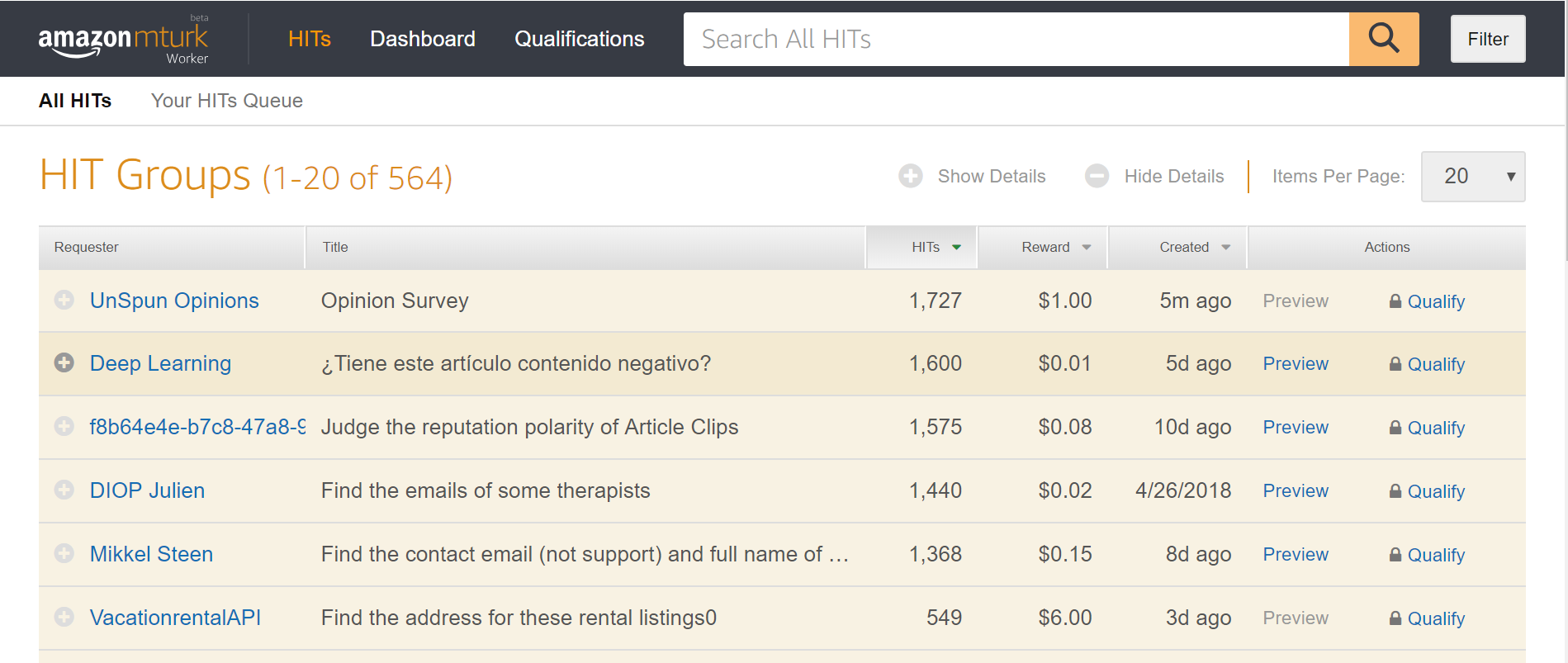}
  \caption{Amazon MTurk HITs Web Portal}
  \label{fig:mturk}
\end{figure} 

\parae{Task Web-UI Portal}
An AMT worker sees HITs listed by the title of the template and the price promised for completing the HIT (\figref{fig:mturk}). (At any given instant, \satyam can be running multiple \satyam-jobs for each supported template). When the worker \textit{accepts} a HIT, she is directed to the \satyam Task Web-UI Portal, which dynamically generates a web page containing one or more \satyam-tasks.  For example, Figure~\ref{fig:tracking_taskpage} shows a Web-UI page for the tracking templates. The generated web page appears as an IFrame within the AMT website. When the worker submits the HIT, the results are entered into the {\it Result-Table} and AMT is notified of the HIT completion. 


When dynamically generating the web page, \satyam needs to determine which \satyam-tasks to present to the worker. Listing HITs only by task portal and by price allows \textit{delayed binding} of a worker to \satyam-tasks. \satyam uses this flexibility to (a) achieve uniform progress on \satyam-tasks and (b) avoid issuing the same task to the same worker. When a worker picks a HIT for template $T$ and price $p$, \satyam selects that \satyam-task with the same $T$ and $p$ which has been worked upon the least (using a random choice to break ties). There is on exception to this \textit{least-worked-on} approach. \satyam may need to selectively finish aggregating a few tasks to gather statistics for dynamic price adjustment (described later). In such instances, the least-worked-on mechanism and randomization is restricted to a smaller subgroup rather than the whole task pool, so that the subgroup completes quickly. To avoid issuing the same task to the same worker, \satyam can determine, from the task table, if the worker has already worked on this task (it may present the same task to multiple workers to improve result quality, \secref{sec:fusion}). A single HIT may contain multiple \satyam-tasks, so \satyam repeats this procedure until enough tasks have been assigned to the HIT.

\parae{Groundtruth Compilation}
Once all the tasks corresponding to a job have been purged (\secref{sec:payments}), this component compiles all the aggregated results corresponding to this job into a JSON file, stores that file at a user-specified location and notifies the user. Before ground-truth compilation, \sysname may need to \textit{post-process} the results. Specifically, for video-based job categories like tracking, \satyam must \textit{stitch} video chunks together to get one seamless groundtruth for the video. We omit the details of the stitching algorithm, but it uses the same techniques as the groundtruth-fusion tracking algorithm (\secref{sec:fusion:details}) to associate elements in one chunk with those in overlapped frames in the next chunk.

\begin{figure*}[t]
\centering
\begin{minipage}{0.65\columnwidth}
  \centering
  \includegraphics[width=0.9\columnwidth]{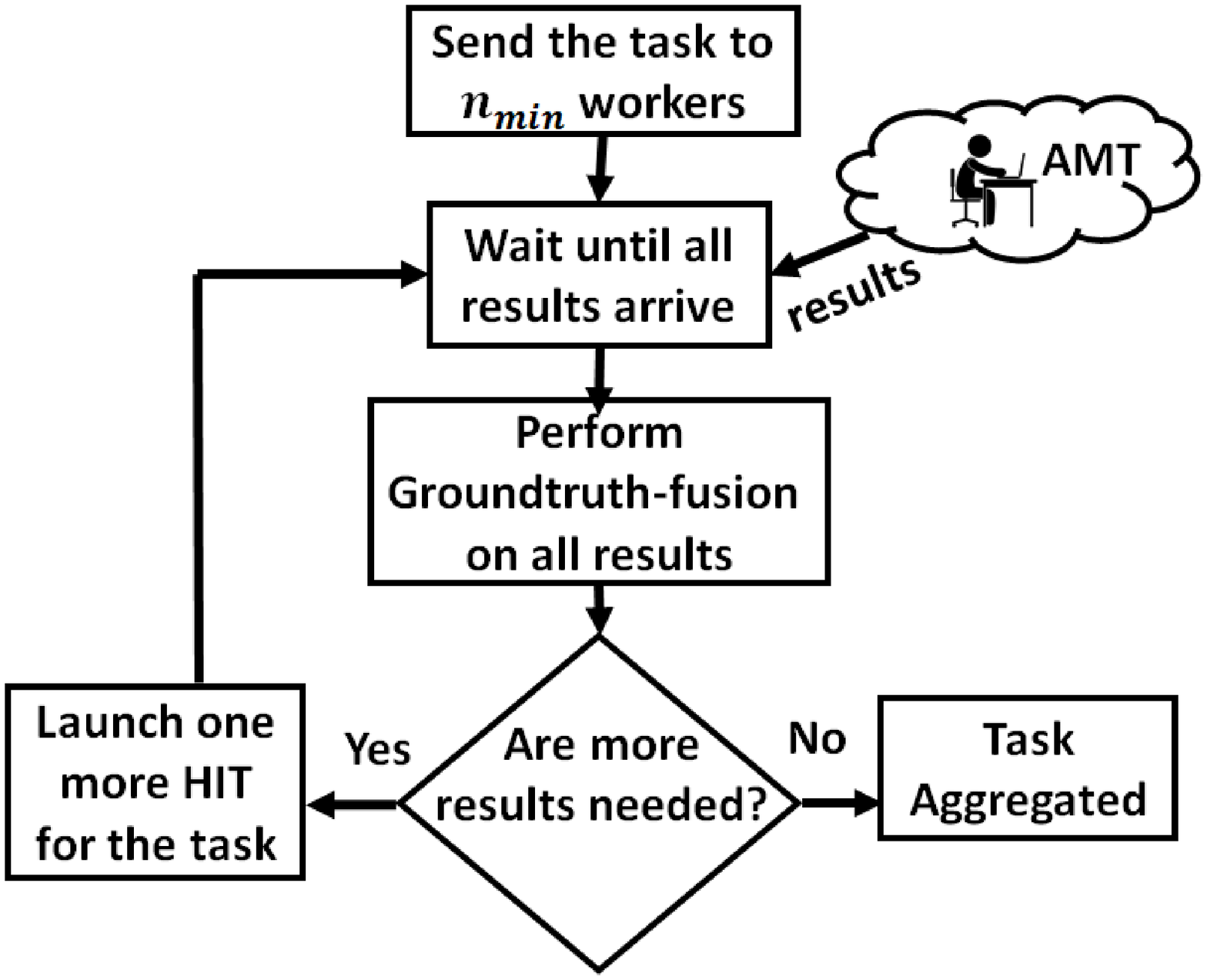}
 \caption{Quality Control Loop in \sysname}
\label{fig:QualityControlLoop}
\end{minipage}
\begin{minipage}{0.6\columnwidth}
  \centering
  \includegraphics[width=0.8\columnwidth]{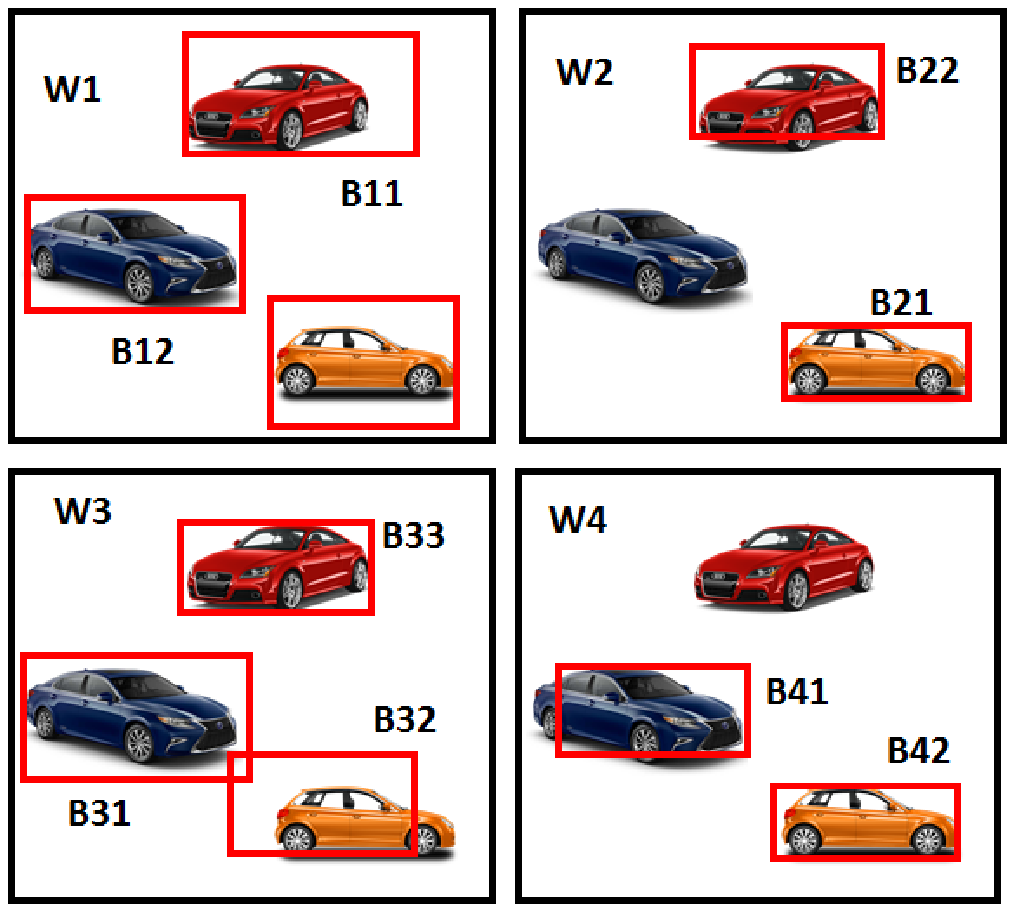}
  \vspace{0.15in}
 \caption{Example groundtruth fusion in Multi-object Detection}
\label{fig:MultiObjectLocalizationFusion}
\end{minipage}
\begin{minipage}{0.65\columnwidth}
  \centering
  \vspace{3mm}
  \includegraphics[width=0.9\columnwidth]{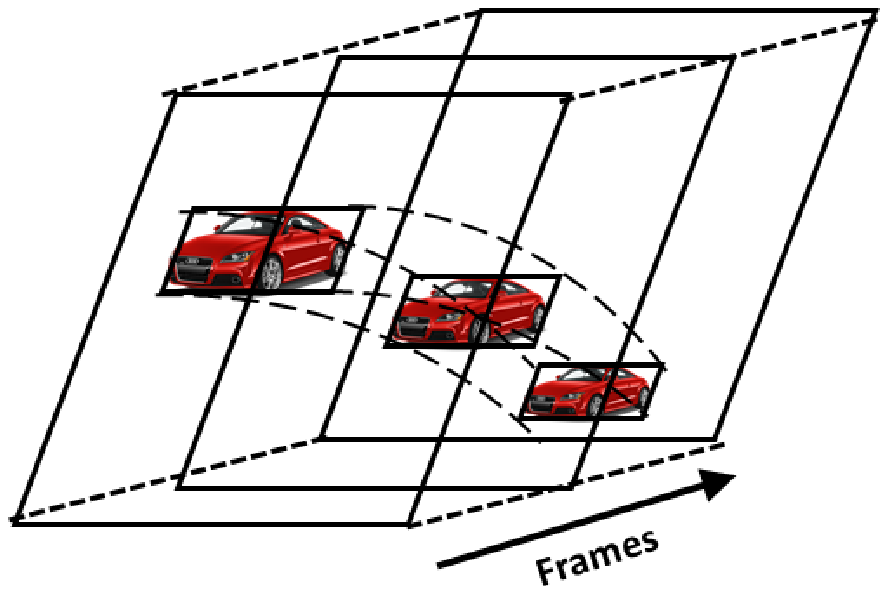}
 \vspace{6mm}
 \caption{Groundtruth fusion in Multi-object Tracking is a 3-D extension of Multi-object detection}
 
\label{fig:MultiObjectTrackingFusion}
\end{minipage}
\end{figure*}

\section{Quality Control}
\label{sec:fusion}

\sysname's quality control relies on the wisdom of the crowds~\cite{wisdom}: {\it when a large enough number of non-colluding workers independently agree on an observation, it must be ``close'' to the groundtruth}. To achieve this \sysname solicits groundtruth for the same image/video clip from multiple workers and only accepts elements of the groundtruth that have been \textit{corroborated} by multiple workers. For instance, in a detection task with several bounding boxes, only those, for which at least 3 workers have drawn similar bounding boxes, are accepted. 

\figref{fig:QualityControlLoop} depicts \satyam's quality control loop. One instance of the loop is applied to each \satyam-task. \sysname first sends the same task to  $n_{min}$ workers to obtain their results. $n_{min}$ depends on the job category and is typically higher for more complex tasks (described in more detail below). In the {\it groundtruth-fusion} step, \satyam attempts to corroborate and fuse each groundtruth element (\eg bounding box) using a job category specific groundtruth-fusion algorithm. If the fraction of corroborated elements in an image/video clip is less than the coverage threshold ($\eta_{cov}$), \sysname determines that more results need to be solicited and relaunches more HITs, one at a time. For some images/videos, even for humans, agreeing on groundtruth maybe difficult. For such tasks we place a maximum limit $n_{max}$ (20 in our current implementation) on the number workers we solicit groundtruth from. The task is marked ``aggregated'' and removed from the task list either if we reach the maximum limit or if the fraction of corroborated elements exceeds $\eta_{cov}$.

\subsection{Dominant Compact Cluster}
\label{sec:fusion:overview}

All the groundtruth-fusion algorithms in \sysname are based on finding the 
{\it Dominant Compact Cluster} (DCC) which represents {\it the set of similar results that the largest number of workers agree on}. If the number of elements in the dominant compact set is greater than $n_{corr}$, the groundtruth for that element is deemed as corroborated. 

\parab{Definition}
Suppose that $n$ workers have generated $n$ versions of the groundtruth $E_1,E_2,...,E_n$ for a particular element in the image/video (as in \figref{fig:MultiObjectLocalizationFusion} where each of the 4 workers has drawn bounding box around the orange car). 
For each job category, we define a distance metric $D(E_i,E_j)$ that is higher the more dissimilar $E_i$ and $E_j$ are. A fusion function $F_{fusion}(E_1,E_2,\cdots,E_k) = E_{fused}$ specifies how different versions of the groundtruths can be combined into one (\eg by averaging multiple bounding boxes into one).  
All {\it groundtruth-fusion} algorithms start by clustering ${E_1,E_2,\cdots,E_n}$ based on $D$ while guaranteeing that none of the elements of its cluster is farther than $\tau$ distance from the fused element \ie $D(E_{fused},E_k) < \tau$ for all $E_k$ within a cluster. $\tau$, the {\it compactness constraint}, ensures that the clusters do not have any results that are too dissimilar from each other. After the clustering, the cluster with the most number of elements is deemed the {\it dominant compact cluster} and  $E_{fused}$ computed over this cluster is deemed the cluster head. \looseness=-1

\parab{Greedy Hierarchical Clustering to find DCC}
Finding DCC is NP-Hard, so we use greedy hierarchical clustering. We start with $n$ clusters, the $i^{th}$  cluster having the one element $E_i$. At each step, two clusters with the closest cluster heads are merged, provided that the merged cluster does not violate the compactness constraint. The clustering stops as soon as clusters cannot be merged any longer.

\parab{Variations across different templates}
While finding the DCC is common across all groundtruth-fusion algorithms, the specific values and functions such as $n_{min}$, $D(E_i, E_j)$, $n_{corr}$, $F_{fusion}$, $\eta_{cov}$ and  $\tau$ are different for each fusion algorithm. In the rest of this section, we describe the various choices we use for these values and functions.

\subsection{Fusion Details}
\label{sec:fusion:details}

\parab{Image and Video Classification}
For this template, \satyam uses a super-majority criterion, selecting that class for which the fraction of workers that agree on the class exceeds $\beta \in (0,1)$ (we chose $\beta = 0.7$, \secref{sec:evaluation:paramChoices}). This is equivalent to the DCC algorithm with the distance function $D = 0$ if two workers choose the same category and $\infty$ if they do not, $\tau = 0$, and $n_{corr} = \beta n$, where $n$ is the number of results. 

\parab{Counting in Images/Videos}
Given $n$ counts, by $n$ workers, our goal is to robustly remove all the outliers and arrive at a reliable count. We use DCC for this, with  $D(C_i,C_j) = |C_i-C_j|$ where $C_i$ and $C_j$ are the counts from the $i^{th}$ and $j^{th}$ workers. $F_{fusion}$ is chosen as the average of all the counts. $\tau = \lfloor(\epsilon C)\rfloor$, where $C$ is the average count of the cluster, \ie two counts are deemed to be similar only if their deviation is less than $\pm \epsilon$ fraction of the average count. We chose $\epsilon = 0.1$ in our implementation. $n_{min}$ and $n_{max}$ are chosen to be 10 and 20 respectively (\secref{sec:evaluation:paramChoices}).

\parab{Object Detection in Images}
To provide intuition into the groundtruth fusion algorithm for this template we use the example in \figref{fig:MultiObjectLocalizationFusion}, where four workers have drawn bounding boxes around cars in an image. The $j^{th}$ bounding box drawn by the $i^{th}$ worker is represented by $B_{ij}$. A worker may not draw bounding boxes for all cars (\eg $W_2$ and $W_4$), and two different workers may draw bounding boxes on the same image in a different order (\eg $W_1$ draws a box around the red car first, but $W_3$ does it last). Furthermore, workers may not draw bounding boxes consistently: $W_3$'s box around the orange car is off-center and box $B_{11}$ is not tightly drawn around the red car. Our fusion algorithm, designed to be robust to these variations.

\parae{Bounding Box Association}
Since different workers might draw boxes in a different order, we first find the correspondence between the boxes drawn by the different workers. In \figref{fig:MultiObjectLocalizationFusion}, this corresponds to grouping the boxes in the three sets {\small $G_1 =  \{B_{11},B_{22},B_{33}\}$, $G_2 =  \{B_{12},B_{31},B_{41}\}$} and, {\small $G_3 =  \{B_{13},B_{21},B_{32},B_{42}\}$} where each set has boxes belonging to the same car. We model this problem as a multipartite matching problem where each partition corresponds to the bounding boxes of a worker, and the goal is to match bounding boxes of each worker for the same car.

To determine the matching, we use a similarity metric, {\it Intersection over Union (IoU)}, between two bounding boxes, which is the ratio of intersection of the two bounding boxes to their union. Since the matching problem is NP-Hard, we use an iterative greedy approach. For a total of $N$ bounding boxes, we start with $N$ sets with one bounding box per set. At each iteration, we merge the two sets with the highest average similarity  while ensuring that a set may have only one bounding box from a partition. The algorithm terminates when there are no more sets that can be merged. In the end, each set corresponds to all the boxes drawn by different workers for one distinct object in the image.



\parae{Applying groundtruth-fusion on each object}
Once we know the set of bounding boxes that correspond to each other, we can use DCC for fusion. 
Let bounding box {\small $B_i = \left<x_i^{tl},y_i^{tl},x_i^{br},y_i^{br}\right>$} where {\small $(x_i^{tl},y_i^{tl})$ and $(x_i^{br},y_i^{br})$} are the top left and bottom right pixel coordinates respectively. We choose {\small $D(B_i,B_j) = \max(|x_i^{tl}-x_j^{tl}|$,$|y_i^{tl}-y_j^{tl}|$,$|x_i^{br}-x_j^{br}|$, $|y_i^{br}-y_j^{br}|)$, $n_{corr} = 3$, $\tau = 15$} (pixels), {\small $n_{cov} = 0.9$, $n_{min}=5$, $n_{max}=20$}. The fusion-function $F_{fusion}$ generates a fused bounding box as the average of each of top-left and bottom-right pixel coordinates of all the bounding boxes being fused.
Thus, in order to be similar, none of the corners of the boundaries must deviate by more than $\tau$ pixels along the $x$ or $y$ axis. The minimum number of workers to corroborate each box is 3 and 90\% of the boxes need to corroborated before the quality control loop terminates. We arrived at these parameters through a sensitivity analysis (\secref{sec:evaluation:paramChoices}).

\parab{Object Segmentation in Images}
The fusion algorithm used for image segmentation is almost identical to that used for multi-object detection except that bounding boxes are replaced by segments: arbitrary collections of pixels. Thus, while associating segments instead of bounding boxes, the IoU metric is computed by considering individual pixels common to the two segments. For $F_{fusion}$, a pixel is included in the fused segment only if it was included in the annotations of at least 3 different workers. We use $\tau = 1/0.3$, $n_{corr}= 3$, $n_{min}=10$, $n_{max}=20$ and $\eta_{cov}=0.9$. 

\parab{Object Tracking in Videos}
Fusion algorithm for multi-object tracking simply extends that used for multi-object detection to determine a fused \textit{bounding volume}, a 3-D extension of bounding box  (as shown in Figure~\ref{fig:MultiObjectTrackingFusion}). We extend the definition of IoU to a bounding volume by computing and summing intersections and unions over each frame, deemed {\it 3D-IoU}. 
For $F_{fusion}$, we average the bounding boxes across users at each frame independently; this is because different workers may start and end the track at different frames.
We use {\small $\tau = 1/0.3$, $n_{corr}= 3$, $n_{min}=5$, $n_{max}=20$} and {\small $\eta_{cov}=0.9$}.

\subsection{Result Evaluation}
\label{sec:fusion:resultEvaluator}

After all the results for a task have been fused, \satyam \textit{approves} and pays or \textit{rejects} each worker's HIT (\secref{sec:payments}). 


For image and video \textit{classification}, \satyam approves all HITs in which the worker's selected class matches that of the aggregated result. When no class label achieves a super-majority (\secref{sec:fusion:details}), it ranks all classes in descending order of the number of workers who selected them, then chooses the minimum number of categories such that the combined number of workers that selected them is a super-majority, and approves all their HITs.
For \textit{counting}, \satyam approves each worker whose counting error is within $\epsilon$ of the fused count (\secref{sec:fusion:details}).
For object \textit{detection, segmentation and tracking}, \satyam approves each worker whose work has contributed to most of the objects in the image/video. Specifically, \satyam approves a worker if the bounding boxes generated by the worker were in more than half of the dominant compact clusters (\secref{sec:fusion:details}) for objects in the image.


\section{HIT management}
\label{sec:payments}

These components manage the interactions between \sysname and AMT such as launching HITs for the tasks, estimating and adapting the price of HITs to match user specifications, filtering under-performing workers, submitting results to the quality control component, and finally, making/rejecting payments for tasks that have completed. 

\parab{HIT Generator}
This component creates HITs in AMT using the web-service API that AMT provides~\cite{mturk_sdk} and associates these HITs with an entry in the {\it HIT-Table} (which also contains pricing metadata, as well as job/task identification). It ensures that every unfinished task in the \satyam-task table has at least one HIT associated with it in AMT. It does this by comparing the number of unfinished tasks in the Task-Table for each GUID and price level against the number of unfinished HITs in the HIT-Table and determines the deficit. Because a single HIT may comprise multiple tasks, \sysname computes the number of extra HITs needed to fill any deficit and launches them. To determine which HITs have been worked on, as soon as a worker submits a HIT, \satyam records this in the HIT-Table.

\parab{HIT Price Adaptation}
Several organizational and state laws require hourly minimum wage payments. Moreover, hourly wages are easier for users to specify. However, payments in AMT are disbursed at the granularity of a HIT. Thus, \satyam must be able to estimate the ``reasonable'' time taken to do a HIT and translate it to price per HIT based on the desired hourly rate. The time taken for a HIT can vary from a few seconds to several minutes and depends on three factors: (a) the type of the template (\eg segmentation tasks take much longer than classification tasks); (b) even within the same template, more complex jobs can take longer \eg scenes with more cars at a busy intersection; (c) finally, different workers work at different rates. 

To estimate HIT completion times, \satyam instruments the web-UIs provided to workers and measures the time taken by the worker on the HIT. As each job progresses, \satyam continuously estimates the median time to HIT completion per job (considering only approved HITs). It uses this value to adjust the price for each future HIT in this particular job. Using this, \satyam's price per HIT converges to conform to hourly minimum wage payments (\secref{sec:eval}).

\parab{\sysname HIT Payments} 
Once a task is aggregated, deserving workers must be paid. \satyam relies on the fusion algorithms to determine whether a result should be \textit{accept}-ed or not (\secref{sec:fusion}). A single HIT may include multiple \satyam-tasks; \satyam's HIT Payments component computes the fraction of \textit{accepted} results in a HIT across all of these tasks and pays the worker if this fraction is above a threshold.

\parab{Worker Filtering}
Worker performance can vary across templates (\eg good at classification but not segmentation), and across jobs within a given template (\eg good for less complex scenes but not for more complex ones). To minimize rejected payments, \satyam tracks worker performance and avoids recruiting them for tasks they might perform poorly at. To do this, as \satyam rejects payments to undeserving workers for a certain task, it tracks worker approval rates (using the AMT-supplied opaque workerID) for each job and does not serve HITs to workers that have low approval rates (lower than 50\% in our implementation). While serving HITs to workers with past high performance history allows \satyam to be efficient, \satyam must also explore and be able to discover new workers. Thus, \satyam allows workers with good approval rates to work on 80\% of the HITs, reserving the rest for workers for whom it does not have any history. As shown in our evaluations~\ref{sec:evaluation:workerFiltering}, worker filtering results in much fewer overall rejections. 



\parab{\sysname Task Purge}
This component, triggered whenever a result is aggregated, removes completed tasks from the Task Table so that they no longer show up in any future HITs.

\section{Evaluation}
\label{sec:eval}

We have implemented all components (\figref{fig:Satyam-Components}) of \satyam on Azure. Our implementation is 13635 lines of C\# code. Using this, we evaluate \sysname by comparing the fidelity of its groundtruth against public \textit{ML benchmark} datasets. In these benchmarks, groundtruth was curated/generated by trained experts or by using specialized equipment in controlled settings. 
To demonstrate \sysname's effectiveness in a real world deployment we generate a data set by extracting images from four video surveillance streams at major traffic intersections in two US cities.

We evaluate \satyam along the following dimensions:
 (a) The quality of ground truth obtained by \sysname compared with that available in popular benchmark data sets;
(b) The \textit{accuracy} of deep neural networks trained using groundtruth obtained by \sysname compared with those trained using benchmark data sets;
 (c) The efficacy of fine-tuning in a deployed real-world vision-based system;
 (d) The \textit{cost} and \textit{time} to obtain groundtruth data using \sysname and; 
 (e) The efficacy of our adaptive pricing and worker filtering algorithms 
(f) The sensitivity of groundtruth-fusion algorithms to parameters.


 \subsection{ML Benchmark Datasets}
 \label{sec:eval:datasets}
  
 \parab{Image Classification (ImageNet-10)} We create this dataset by picking all the images corresponding to 10 classes commonly seen in video surveillance cameras from the ImageNet~\cite{ImageNet} dataset. Our dataset contains 12,482 images covering these classes: cat, dog, bicycle, lorry-truck, motorcycle, SUV, van, female person and male person.  

 \parab{Video Classification (JHMDB-10)} For this data set we pick all the video clips from the JHMDB~\cite{JHMDB} data set corresponding to to 10 common human activities: clap, jump, pick, push, run, sit, stand, throw, walk and, wave (a total of 411 video clips).

 \parab{Counting in Images (CAPRK-1)} We create this data set by selecting 164 drone images taken from one parking lot from CAPRK~\cite{carpk} (a total of 3,616 cars).

 \parab{Object Detection in Images (KITTI-Object)}
 We create this data set by considering 3 out of 8 classes (cars, pedestrians and cyclists) in the KITTI~\cite{KITTI}  data set with 8000 images (a total of 20,174 objects.). The groundtruth in KITTI established using LiDAR mounted on the car. 

 \parab{Object Segmentation in Images (PASCAL-VOC-Seg)} PASCAL-VOC \cite{pascalvoc} is a standardized image dataset for object classification, detection, segmentation, action classification, and person layout. 
 We create this data set by choosing 353 images from the PASCAL-VOC \cite{pascalvoc} data set that have segmentation labels, including the groundtruth of both class- and instance-level segmentation, corresponding to a total of 841 objects of 20 different classes.

 \parab{Tracking in Videos (KITTI-Trac)} For this dataset we chose all 21 video clips that were collected from a moving car from KITTI~\cite{KITTI} (about 8000 frames), but evaluate tracks only for 2 classes -- cars and pedestrians. During the pre-processing step, these 21 video clips were broken into 276 chunks of length 3 seconds each with a 0.5 second overlap between consecutive chunks.

 \parab{Traffic Surveillance Video Stream Data (SURV)} We extracted images at 1~frame/minute from the video streams of 4 live HD quality traffic surveillance cameras, over one week (7 days) between 7:00~am and 7:00~pm each day. These cameras are located at major intersections in two U.S cities. We label the dataset corresponding to each of the four cameras as SURV-1, SURV-2, SURV-3 and SURV-4 respectively.    


\begin{figure*}[t]
\centering
  \begin{minipage}{0.66\columnwidth}
  \centering
  \includegraphics[width=0.95\columnwidth]{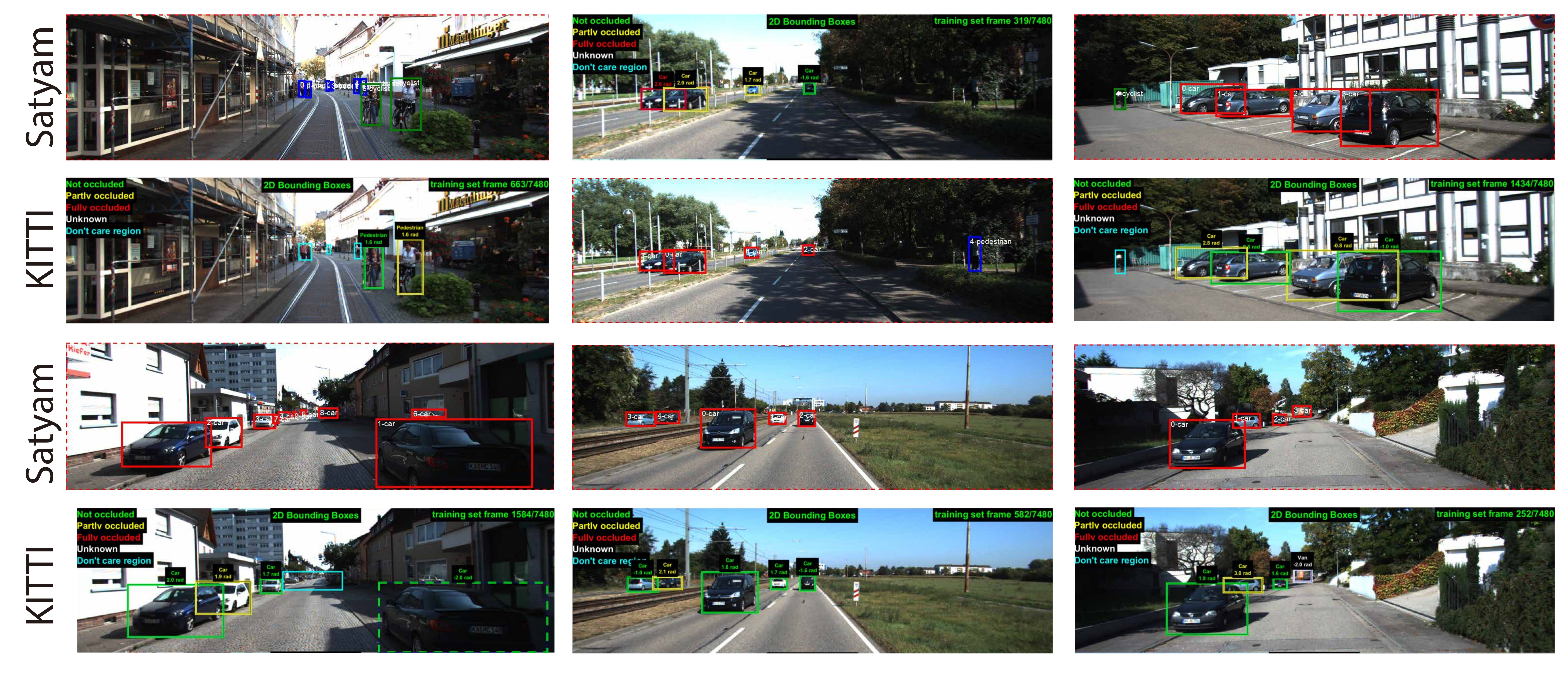}
  \caption{Example Results of \satyam Detection from KITTI}
  \label{fig:DetResult}
 \end{minipage}
\begin{minipage}{0.66\columnwidth}
  \centering \includegraphics[width=0.95\columnwidth]{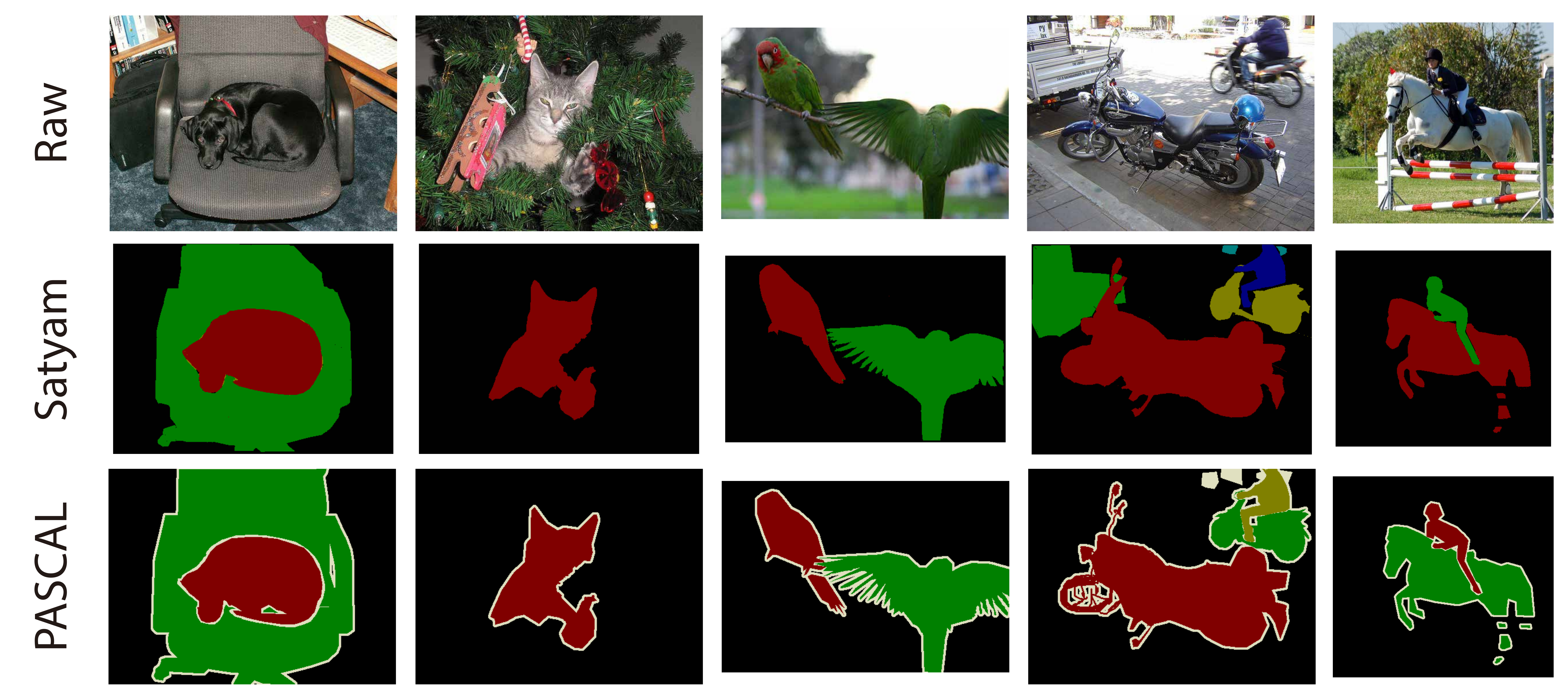}
  \caption{Example Results of \satyam Segmentation from PASCAL}
  \label{fig:SegResult}
\end{minipage}
\begin{minipage}{0.66\columnwidth}
 \centering\includegraphics[width=0.95\columnwidth]{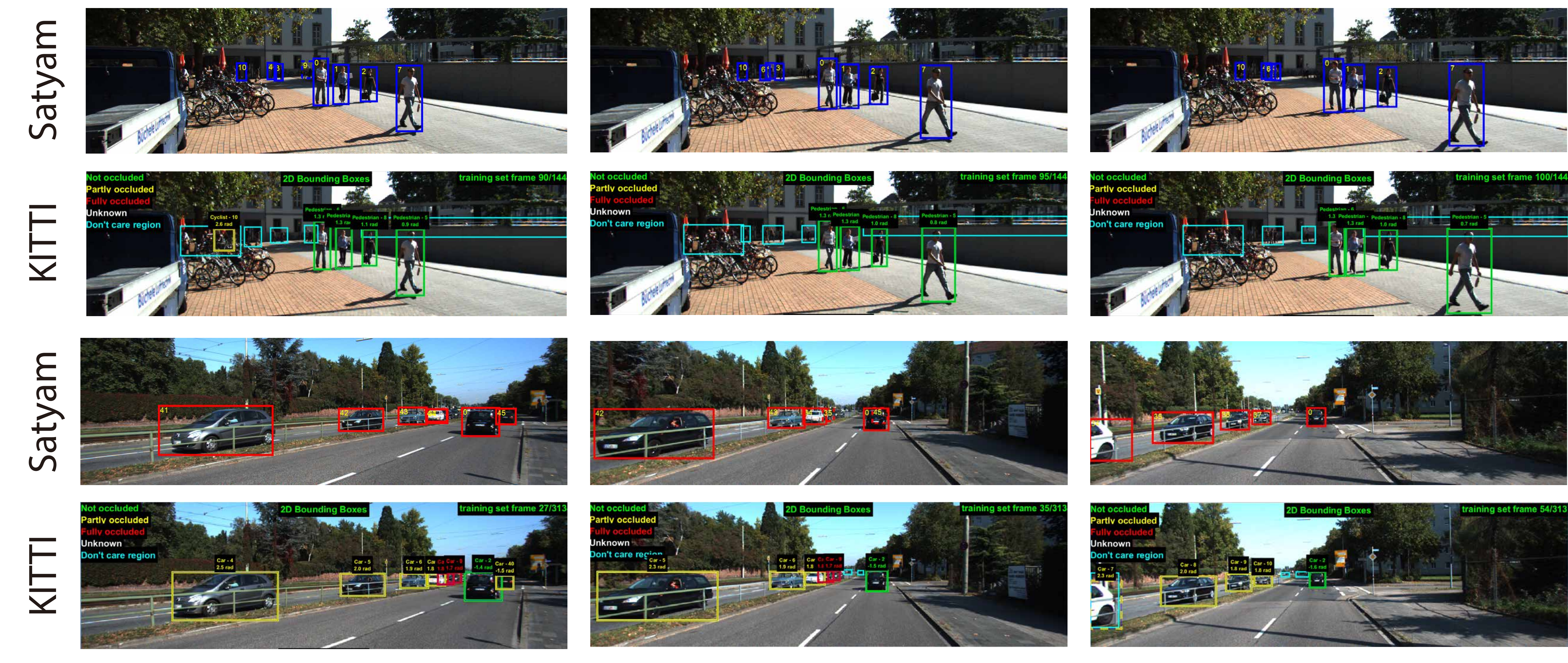}
  \caption{Example Results of \satyam Tracking from KITTI}
  \label{fig:TrackingResult}
\end{minipage}
\end{figure*}

\subsection{Quality of \sysname Groundtruth}
\label{sec:evaluation:accuracy}



\begin{figure}
    \centering
     \includegraphics[width=\columnwidth]{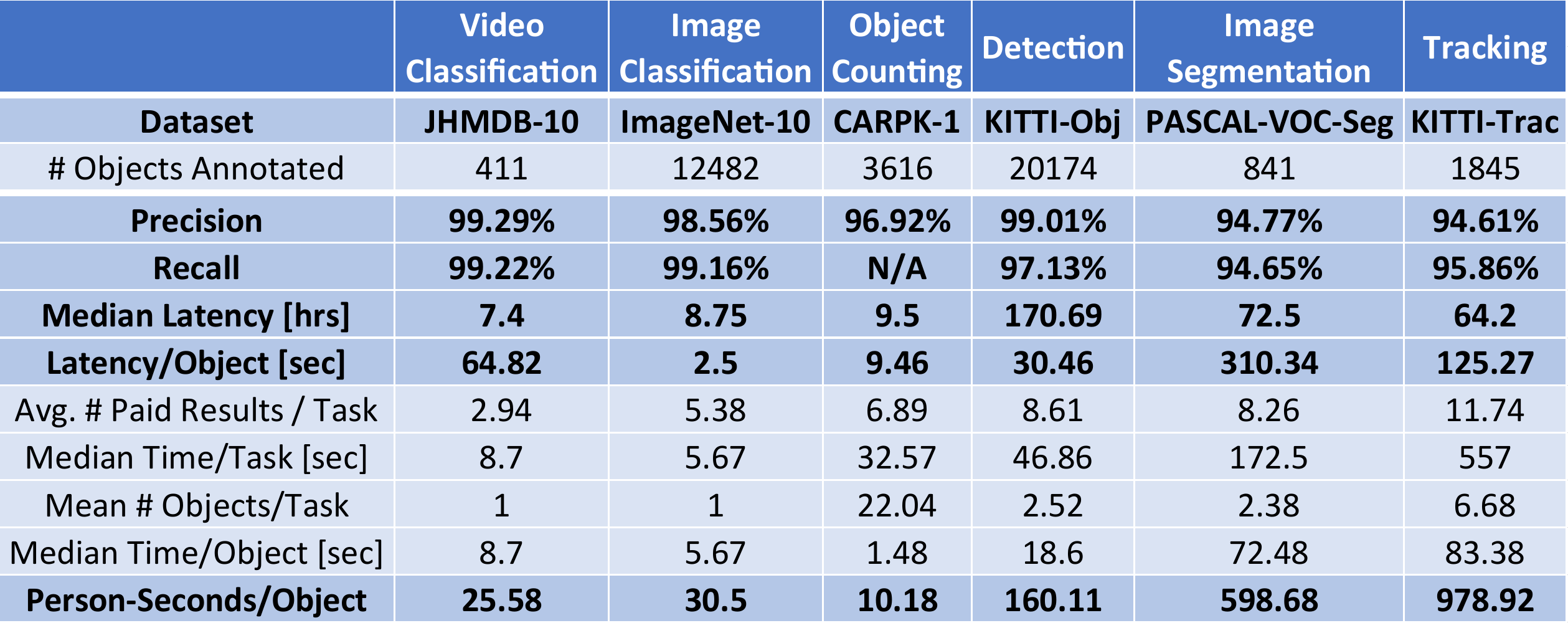}
  \caption{\satyam Accuracy, Latency, and Cost}
  \label{fig:accuracy}
\end{figure}

To demonstrate that \satyam groundtruth is comparable to that in the ML benchmarks, we launched a job in \satyam for each of the six benchmark data sets described in \figref{fig:accuracy}. Figures~\ref{fig:DetResult},~\ref{fig:SegResult} and~\ref{fig:TrackingResult} show some examples of groundtruth obtained using \satyam for detection, segmentation and tracking templates respectively. For this comparison, we evaluate \textit{match-precision} (the degree to which \satyam's groundtruth matches that of the benchmark) and \textit{match-recall} (the degree to which \satyam's workers identify groundtruth elements in the benchmark).

Figure~\ref{fig:accuracy} summarizes \sysname's accuracy for the various templates relative to the benchmarks. \satyam has uniformly high match-precision (95-99\%) and high match-recall ($>$95\%) for the relevant benchmarks. We find that \satyam often deviates from the benchmark because \textit{there are fundamental limits achieving accuracy with respect to popular benchmark data sets}, for two reasons. First, some of the benchmarks were annotated/curated by human experts and have a small fraction of errors or ambiguous annotations themselves. Some of the ambiguity, especially in classification, arises from linguistic confusion between class labels (\eg distinguishing between van and truck). Second, in others that were generated using specialized equipment (\eg LiDAR), part of the generated groundtruth is not perceivable to human eye itself. In the rest of this section, we describe our methodology for each job category and elaborate on these fundamental limits.


\begin{figure}
    \centering
    \includegraphics[width=0.45\columnwidth]{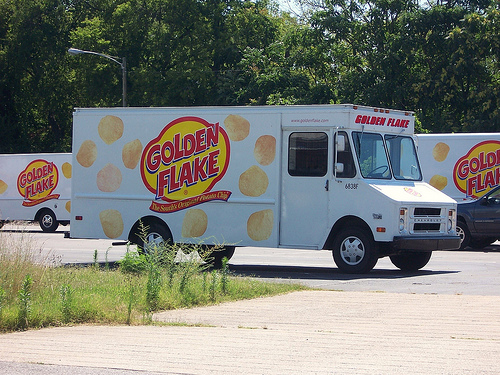}
    \includegraphics[width=0.45\columnwidth]{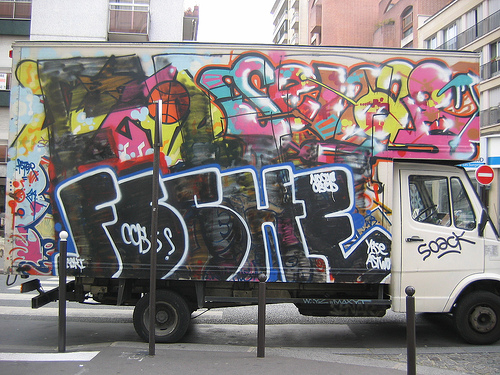}
    \caption{Linguistic confusion between van and truck}
    \label{fig:vantruck1}
\end{figure}

\begin{figure*}
\centering
\begin{minipage}{0.66\columnwidth}
    \centering
    \includegraphics[width=0.9\columnwidth]{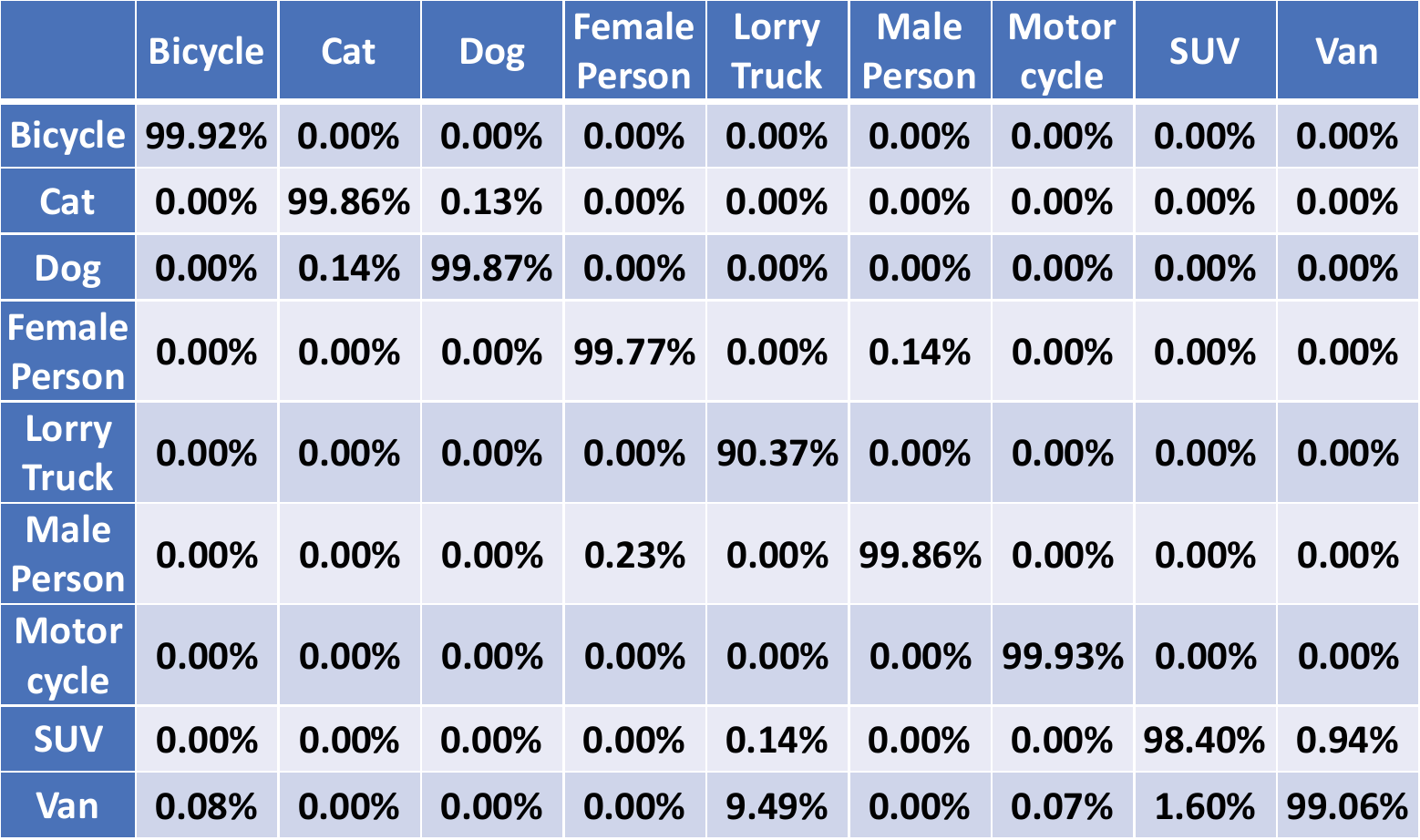}
 \caption{Confusion Matrix of \satyam Result on ImageNet-10}
  \label{fig:confMat}
\end{minipage}
\begin{minipage}{0.66\columnwidth}
    \centering
     \includegraphics[width=\columnwidth]{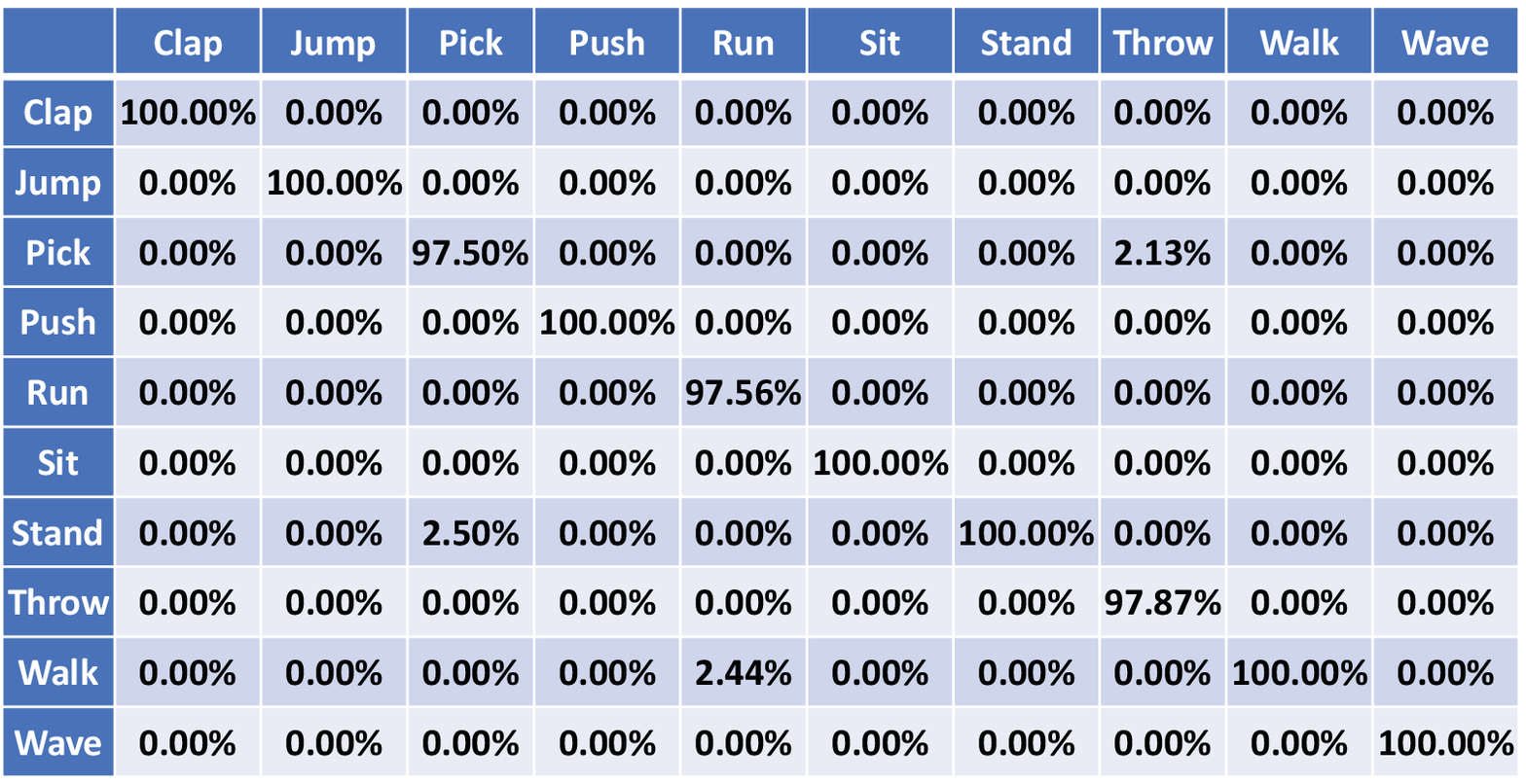}
 \caption{Confusion Matrix of \satyam Result on JHMDB}
 \label{fig:confMatVideo}
\end{minipage}
\begin{minipage}{0.66\columnwidth}
    \centering
    \includegraphics[width=0.95\columnwidth]{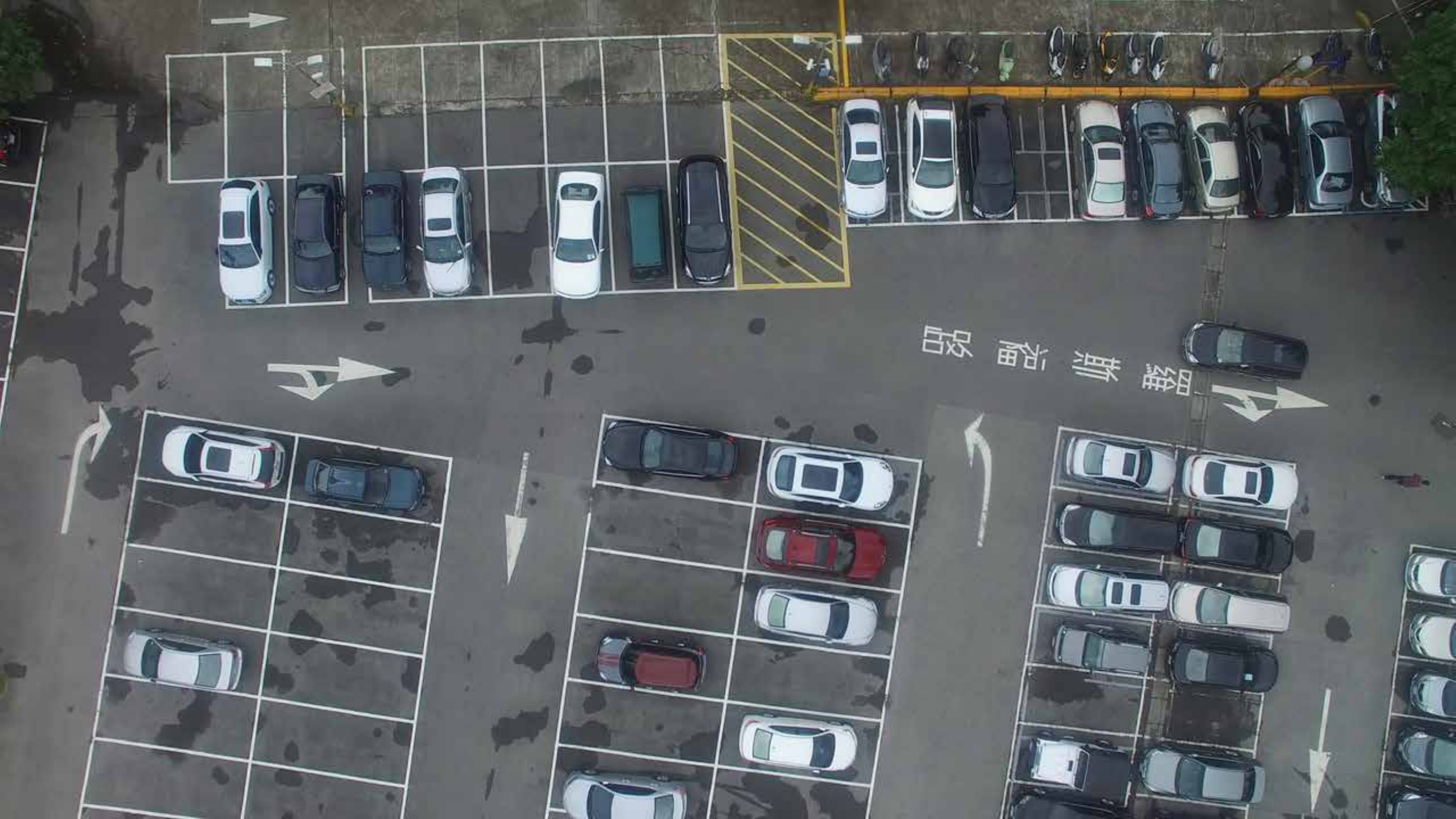}
    \caption{Example of counting error resulting from partially visible cars}
    \label{fig:count_error}
\end{minipage}
\end{figure*}

\parab{Image Classification}
\sysname groundtruth for ImageNet-10 has a match-precision of 98.5\% and a match-recall of 99.1\%. The confusion matrix (\figref{fig:confMat}) for all the 10 categories in ImageNet-10, shows that the largest source of mismatch is from 10\% of vans in ImageNet being classified as lorry-trucks by \satyam. We found that all of vehicles categorized as vans in ImageNet are in fact food or delivery trucks (\eg \figref{fig:vantruck1}), indicating \textit{linguistic confusion} on the part of workers. The only other significant off-diagonal entry in \figref{fig:confMat} at 1.6\% results from linguistic confusion between Vans and SUVs. Discounting these two sources of error, \satyam matches 99.9\% of the groundtruth.

\parab{Video Classification}
\satyam's groundtruth for this category set has a match-precision and match recall exceeding 99\%. The confusion matrix (\figref{fig:confMatVideo}) for the 10 categories in JHMDB-10 reveals only 3 mismatches compared to the benchmark groundtruth. We examined each case, and found that the errors resulted from class label confusion or from incorrectly labeled groundtruth in the benchmark:
a person picking up his shoes was labeled as \textit{standing} instead of \textit{picking}; a person moving fast to catch a taxi was labeled as \textit{walking} instead of \textit{running}; and finally a person who was picking up garbage bags and throwing them into a garbage truck was labeled as \textit{picking up} in JHMDB-10, while \satyam's label was \textit{throwing}. Discounting these cases, \satyam matches 100\% with the groundtruth.

\parab{Counting in Images}
\satyam's car counts deviate from the CAPRK-I benchmark's count groundtruths by 3\% (\figref{fig:accuracy}), which corresponds to an error of 1 car in a parking lot with 30 cars. This arises because of cars that are only partially visible in the image (\eg \figref{fig:count_error}), and workers were unsure whether to include these cars in the count or not. By inspecting the images we found that between 3 and 10\% of the cars in each image were partially visible.


\parab{Object Detection in Images}
For quantifying the accuracy for this template, we adopt the methodology recommended by the KiTTI benchmark -- two bounding boxes are said to match if their IoU is higher than a threshold. \sysname has a high match-precision of 99\% and match-recall of 97\% (\figref{fig:accuracy}). The match-recall is expected to be lower than match-precision: the LiDAR mounted on KITTI's data collection vehicle can sometimes detect objects that may not be visible to the human eye.

\begin{figure}
    \centering
    \includegraphics[width=0.32\columnwidth]{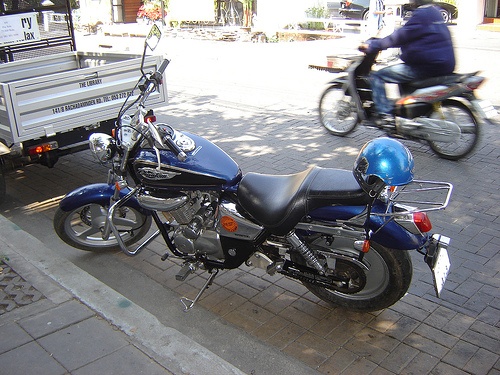}
    \includegraphics[width=0.32\columnwidth]{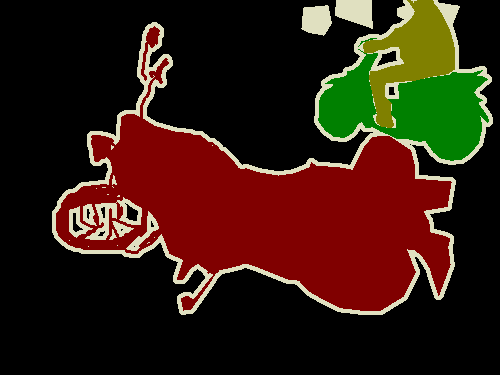}
    \includegraphics[width=0.32\columnwidth]{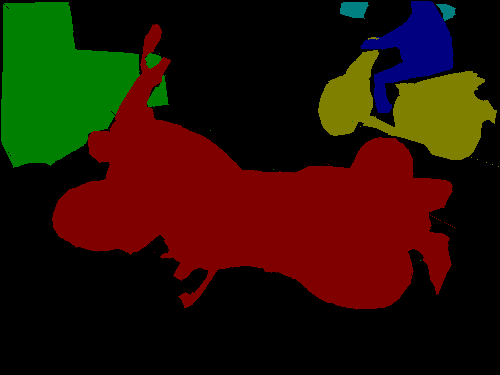}
    \caption{Example of missing segmentation labels from PASCAL. From left to right: raw image, PASCAL label, \satyam label. \satyam segments a small truck on the top left corner which was not present in the ground truth.}
    \label{fig:segmentation_error}
\end{figure}

\parab{Object Segmentation in Images} We use Average Precision (AP)~\cite{pascalvoc} to quantify the accuracy. We use a range of IoUs (0.5-0.95 with steps of 0.05) to compute the average to avoid a bias towards a specific value. \satyam achieves an AP of 90.03\%. We also provide a match-precision of 94.77\% and a match-recall of 94.56\% using an IoU of 0.5. The dominant cause of false positives is missing annotations in the ground-truth. \figref{fig:segmentation_error} shows examples of such missing annotations from PASCAL that \satyam's users were able to produce. The primary cause of false negatives is that our experiments used a lower value of $\eta_{cov}$ than appropriate for this task; we are rectifying this currently.

\parab{Object Tracking in Videos}
A track is a sequence of bounding boxes across multiple frames. Consequently,  we use the same match criterion for this template as detection across all the video frames. As seen from (\figref{fig:accuracy}), \sysname has a match-precision and match-recall of around 95\%. To understand why, we explored worker performance at different positions in the chunk: we found that, as workers get to the end of a chunk, they tend not to start tracking new objects. Decreasing the chunk size and increasing the overlap among consecutive chunks would increase accuracy, at higher cost. 



\begin{figure*}[t]
  \centering
  \begin{minipage}{0.66\columnwidth}
    \centering
    \includegraphics[width=0.95\columnwidth]{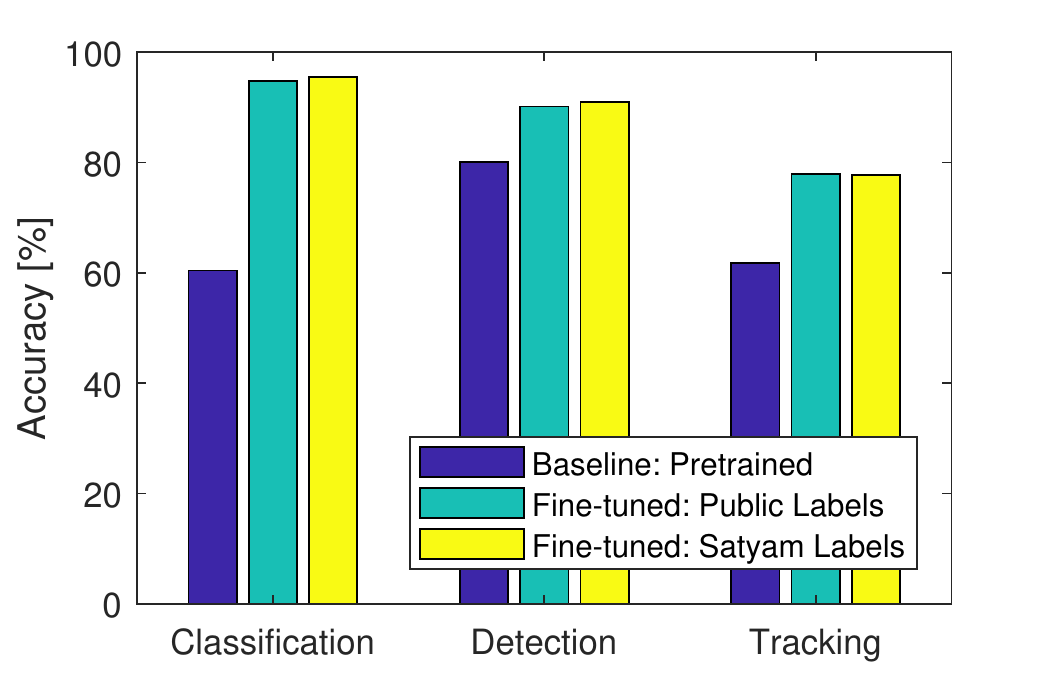}
  \caption{Training Performance of \satyam}
  \label{fig:TrainingAccuracy}
  \end{minipage}
  \begin{minipage}{0.66\columnwidth}
    \centering
  \includegraphics[width=0.95\columnwidth]{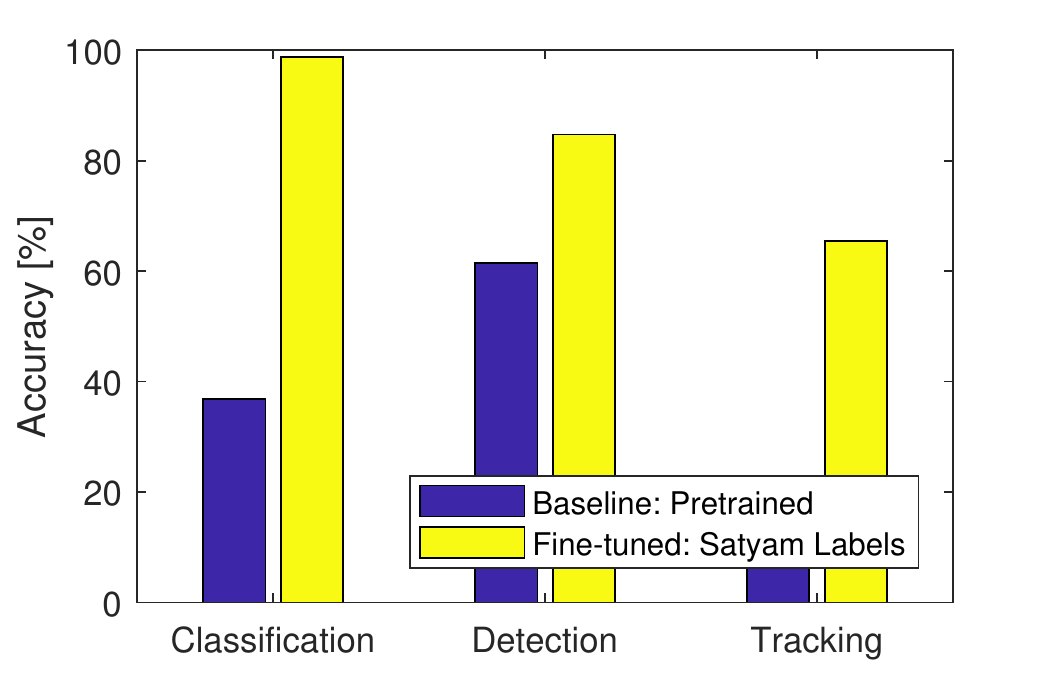}
  \caption{End to End Training using \satyam Labels}
  \label{fig:E2ETrainingAccuracy}
  \end{minipage}
 \begin{minipage}{0.66\columnwidth}
  \centering
  \includegraphics[width=0.9\columnwidth]{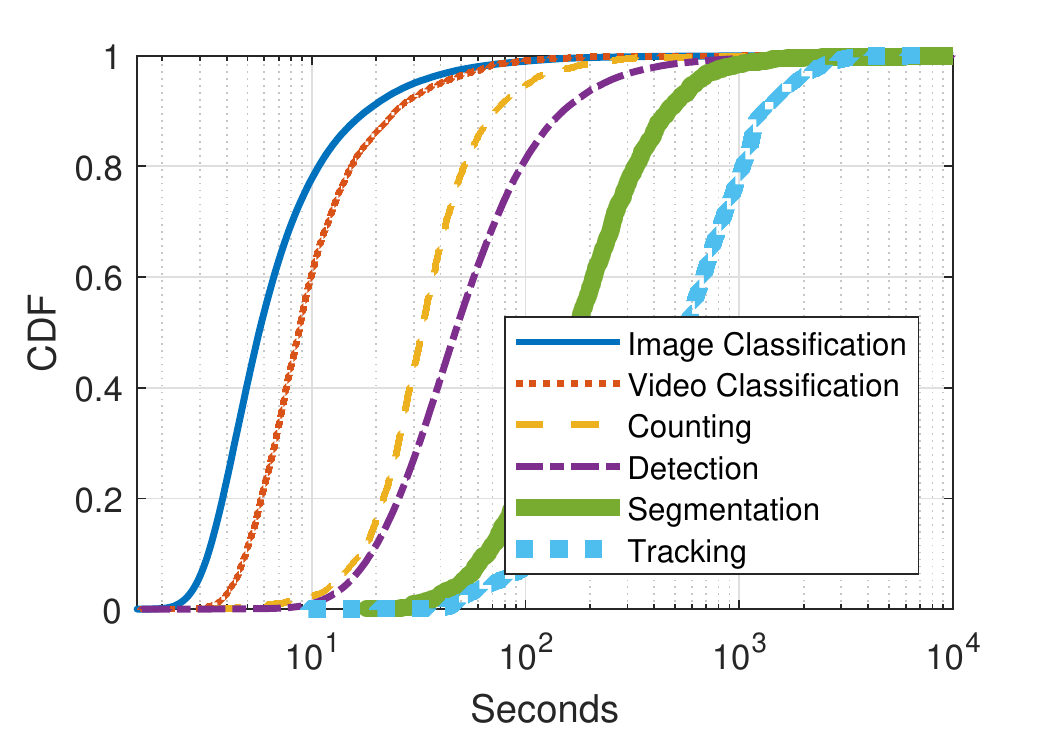}
  \caption{CDF of Time Spent Per Task of All Job Categories}
  \label{fig:timeCDF}
\end{minipage}
\end{figure*}

\subsection{Re-training Models using \sysname}
\label{sec:evaluation:trainingPerformance}

A common use case for \satyam is fine-tuning ML models for improving their performance using data specific to a deployment (similar to example in Section~\ref{sec:intro}). In this section, we validate this observation by showing that (\figref{fig:TrainingAccuracy}): a) re-training ML models using \satyam groundtruth outperforms off-the-shelf pre-trained models, and b) models retrained using \satyam groundtruth perform comparably with models retrained using benchmarks. When re-training and testing a model, either with \satyam or benchmark groundtruth, we use standard methodology to train on 80\% of the data, and test on 20\%. In all cases, we retrain the last layer using accepted methodology~\cite{retrain1,retrain2}.

\parab{Image Classification}
For this job category, we evaluate retraining a well-known state-of-the-art image classification neural network, Inception (V3) \cite{inception}. The original model was pre-trained on ImageNet-1000 \cite{ImageNet} for 1000 different classes of objects. Using this model as-is on ImageNet-10 yields a classification accuracy (F1-score~\cite{F1score}) of about 60\% (\figref{fig:TrainingAccuracy}). Retraining the models using the ImageNet-10 groundtruth increases their accuracy to 94.76\%, while retraining on \satyam results in an accuracy of 95.46\%.



\parab{Object Detection in Images}
For this category, we evaluate YOLO~\cite{yolo}, pre-trained on the MS-COCO dataset. Our measure of accuracy is the mean average precision~\cite{meanAP}, a standard metric for object detection and  localization that combines precision and recall by averaging precision over all recall values. The pre-trained YOLO model has high (80\%) mean average precision, but retraining it using KITTI-Object increases this to 90.1\%. Retraining YOLO using \satyam groundtruth matches KITTI-Object's performance, with a mean average precision of 91.0\% (\figref{fig:TrainingAccuracy}).

\parab{Tracking in Videos}
As of this writing, the highest ranked open-source tracker on the KITTI Tracking Benchmark leaderboard is MDP~\cite{MDP}, so we evaluate this tracker (with YOLO as the underlying object detector) using the standard Multi-Object Tracking Accuracy (MOTA~\cite{MOT_Metric}) metric, which also combines precision and recall. MDP using YOLO-CoCo's detections achieves a MOTA of 61.83\% as depicted in \figref{fig:TrainingAccuracy} but fine-tuning Yolo's last layer using the labels from KITTI and \satyam improve MOTA to  78\% and 77.77\% respectively. Further investigation reveals that the improvement in MoTA from fine-tuning was primarily due to improvement in recall -- while precision was already high (98\%) before fine-tuning, recall was only 63.54\%. After fine-tuning recall improved to 81.70\% for KITTI and 83.24\% for \sysname.



\begin{figure}
   \centering\includegraphics[width=0.8\columnwidth]{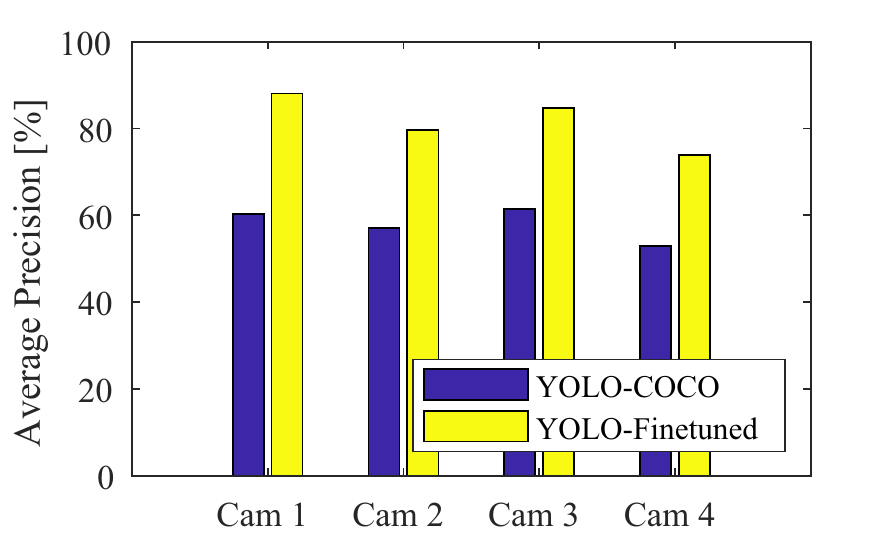}
  \caption{Improvement of performance with fine-tuned YOLO.}
  \label{fig:MotivationYoLoPerformance}
\end{figure}


\begin{figure*}
\centering
\begin{minipage}{0.66\columnwidth}
  \centering
\includegraphics[width=\columnwidth]{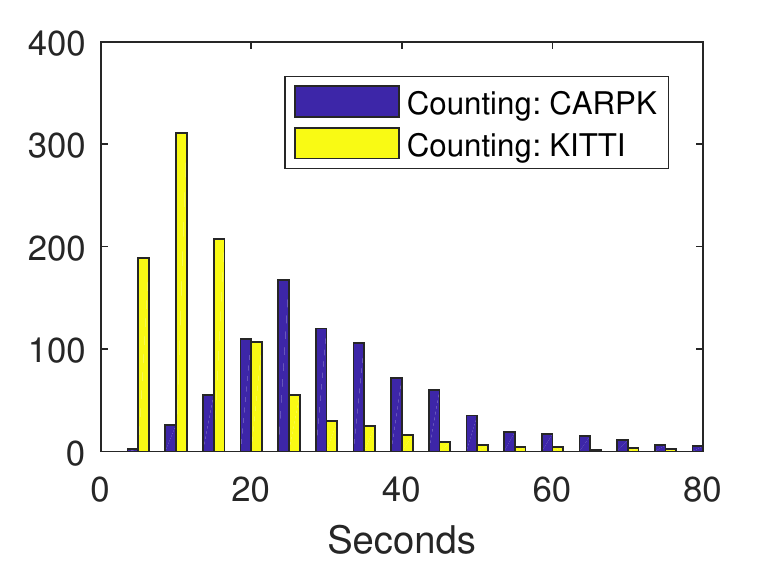}
  \vspace*{-0.2in}
  \caption{Histogram of Time Per Counting Task over Different Datasets}
  \label{fig:timeCDF_counting}
\end{minipage}
\begin{minipage}{0.66\columnwidth}
  \centering
  \includegraphics[width=\columnwidth]{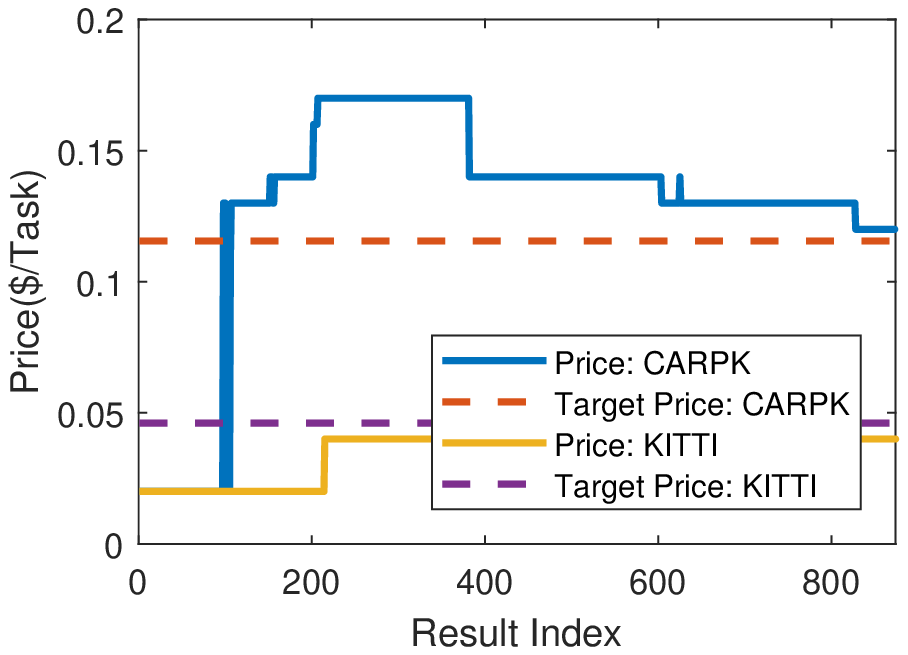}
  \vspace*{-0.2in}
  \caption{Adaptive Pricing on Counting Task}
  \label{fig:adaptivePricing}
\end{minipage}
\begin{minipage}{0.66\columnwidth}
\centering
  \vspace{4mm}
  \includegraphics[width=0.95\columnwidth]{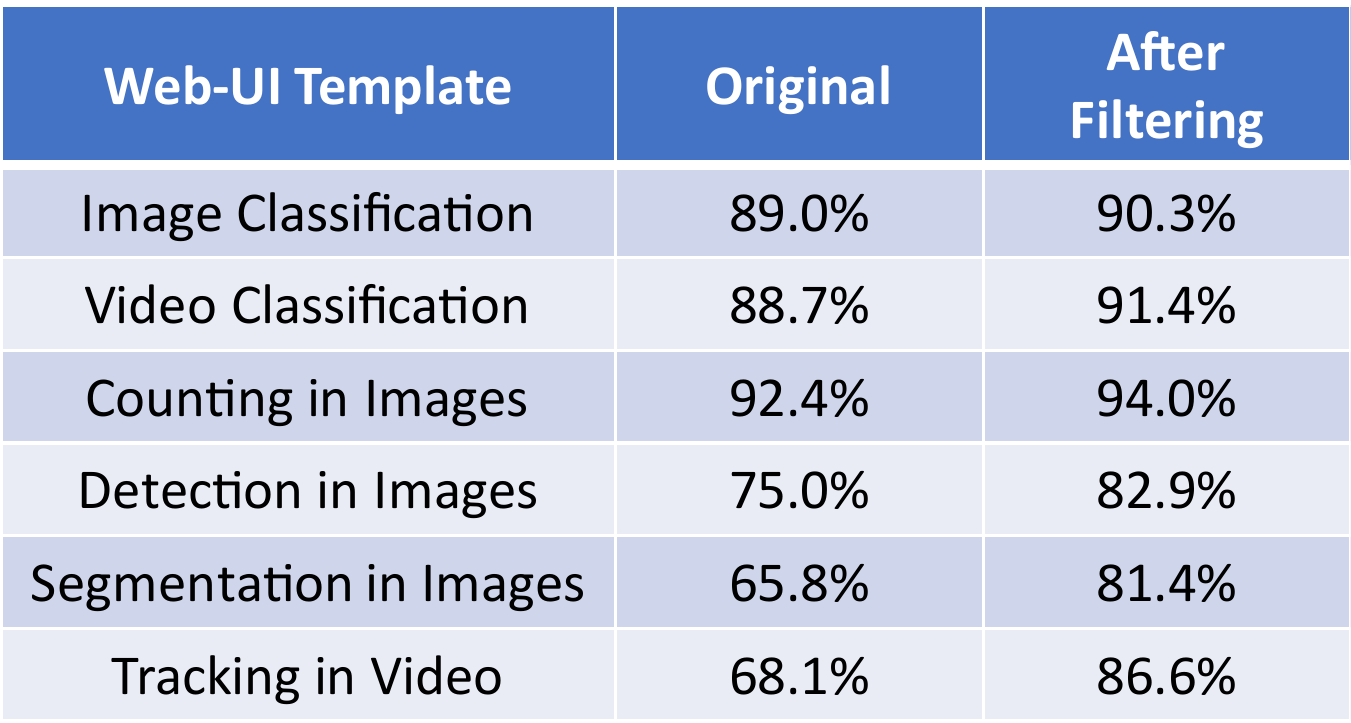}
  \vspace{4mm}
  \caption{Approval Rates for various \sysname templates}
\label{tab:approvals}
\end{minipage}
\end{figure*}

\subsection{\sysname In Real-World Deployments}
\label{sec:evaluation:SURVData}

In order to evaluate the impact of using \satyam in the real world, we extracted images at 1~frame/minute from the video streams of 4 live HD quality traffic surveillance cameras (labeled SURV-1 to SURV-4), over one week (7 days) between 7:00~am and 7:00~pm each day. These cameras are located at major intersections in two U.S cities.

We now show that using \satyam groundtruth to fine-tune ML models can result in improved classification, detection, and tracking performance. For this, we use the SURV dataset, which has surveillance camera images from four intersections, to obtain ground-truth with \satyam, then re-trained YOLO-CoCo~\cite{retrain1,retrain2} with 80\% of the ground-truth and tested on the remaining 20\%.

\satyam re-training can improve YOLO-CoCo performance uniformly across the four surveillance cameras (\figref{fig:MotivationYoLoPerformance}). The average precision improves from 52-61\% for the pre-trained models to 73-88\% for the fine-tuned models -- an improvement of 20-28\%. This validates our assertion that camera fine-tuning will be essential for practical deployments, motivating the need for a system like \satyam. 

\figref{fig:E2ETrainingAccuracy} demonstrates that these benefits carry over to other job categories as well and shows fine-tuning Inception v3 for classification for one of the cameras, SURV-3. To compute this result, we used our groundtruth data from SURV-3 for the detection task, where workers also labeled objects, then trained Inception to focus on one object type, namely cars. While the pre-trained Inception model works poorly on SURV-3, fine-tuning the model results in an almost perfect classifier. Similarly, fine-tuning also results in an almost 40\% improvement in the MOTA metric for the tracker. 


\begin{figure*}
\centering
\begin{minipage}{0.66\columnwidth}
  \centering
\includegraphics[width=\columnwidth]{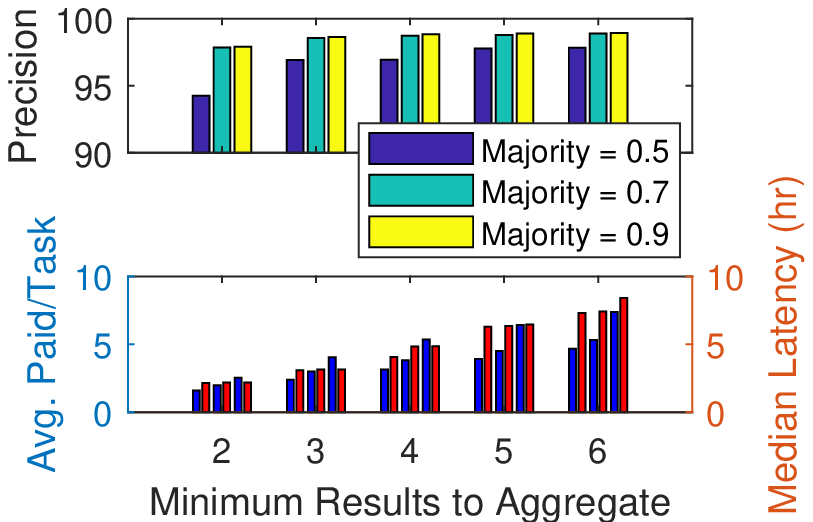}
  \caption{Accuracy, Latency and Cost: Image Classification}
  \label{fig:param:image_classification}
\end{minipage}
\begin{minipage}{0.66\columnwidth}
  \centering
  \includegraphics[width=\columnwidth]{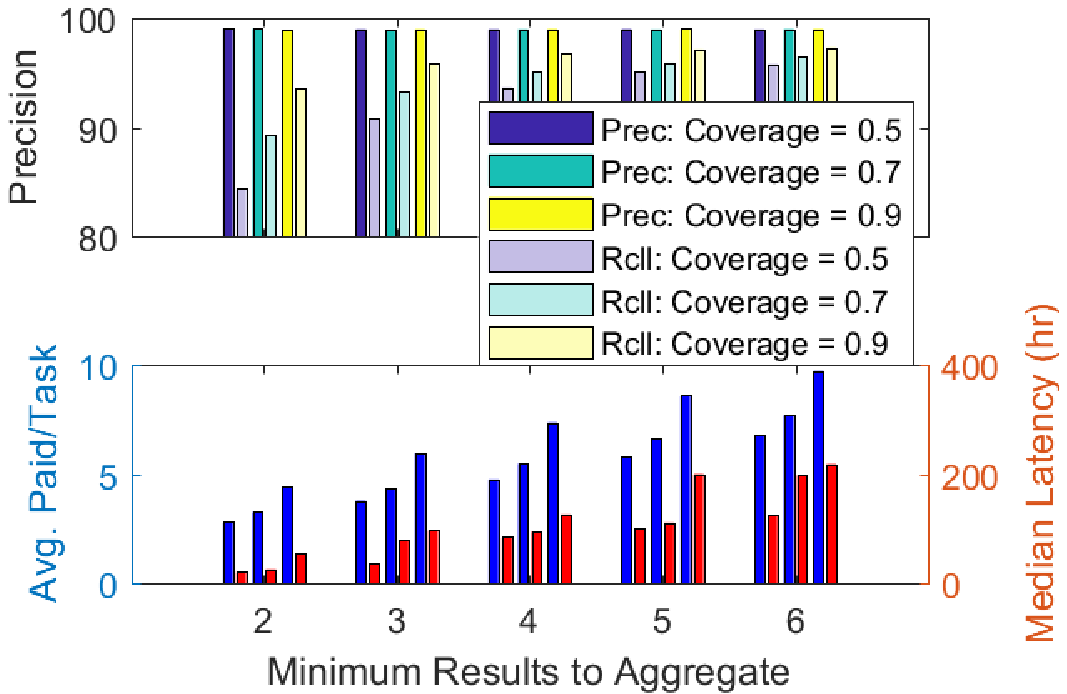}
  \caption{Accuracy, Latency and Cost: Detection}
  \label{fig:param:localization}
\end{minipage}
\begin{minipage}{0.66\columnwidth}
\centering
 \includegraphics[width=\columnwidth]{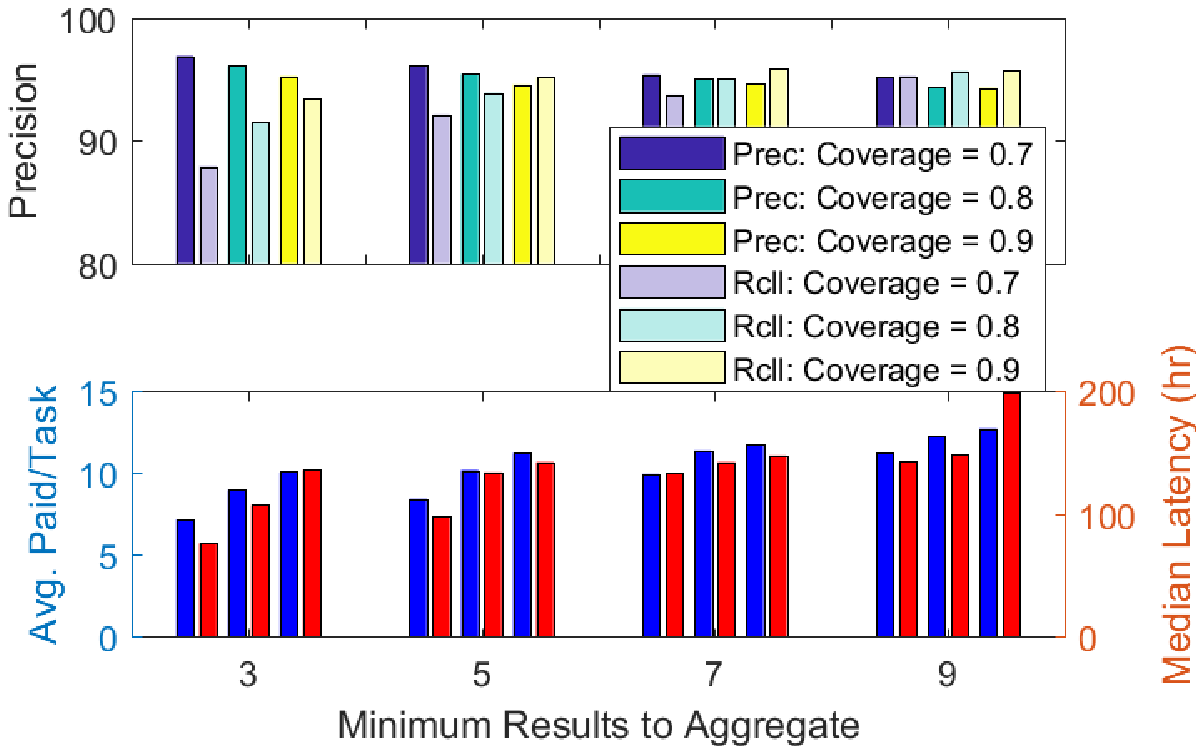}
  \caption{Accuracy, Latency and Cost: Tracking}
  \label{fig:param:tracking}
\end{minipage}
\end{figure*}

\subsection{Time-to-Completion and Cost}
\label{sec:evaluation:latency_cost}

\figref{fig:accuracy} also shows the median time to complete an entire job, which ranges from 7 hours to 7 days. From this, we can derive the median latency per object, which  ranges from 2.5 seconds/image for image classification to 125 seconds/object for tracking. That figure also shows the cost of annotating an object in \textit{person-seconds/object}: the actual dollar figure paid is proportional to this (\secref{sec:evaluation:priceAdaptation}). By this metric (\figref{fig:accuracy}), image and video classification, and counting cost few tens of person-seconds/object while detection,  segmentation, and tracking require 160,  599, and 978 person-seconds respectively.

\subsection{Price Adaptation}
\label{sec:evaluation:priceAdaptation}

Figure~\ref{fig:timeCDF} is a CDF of the times taken by workers for all the various job categories in our evaluation. It clearly shows that the time taken to complete a task can vary by 3 orders of magnitude across our job categories.
Figure~\ref{fig:timeCDF_counting} depicts the pdf of the times taken for the same category -- counting task --  but for two different data sets \ie CARPK-1 and KITTI-Obj. 
KITTI-Obj has around 10 vehicles on average, and CARPK around 45 in each image, and the distribution of worker task completion times varies significantly across these datasets. (As an aside, both these figures have a long tail: we have seen several cases where workers start tasks but sometimes finish it hours later).

These differences motivate price adaptation. To demonstrate price adaptation in \satyam, we show the temporal evolution of price per HIT for CARPK-1 and KITTI-Obj in Figure~\ref{fig:adaptivePricing}. HIT price for KITTI-Obj converges within 200 results to the ideal target value (corresponding to median task completion time). CARPK-1 convergence is slightly slower due to its larger variability in task completion times (\figref{fig:timeCDF_counting}).


\subsection{Worker Filtering}
\label{sec:evaluation:workerFiltering}

To evaluate the efficacy of worker filtering ran \satyam with and without worker filtering turned on for each of the templates. \figref{tab:approvals} shows that for classification, counting and detection the approval rate is already quite high ranging close to 90\% and thus worker filtering brings about a modest increase in approval rates. For more involved tasks such as tracking and segmentation, the approval rates show a dramatic increase from 60\% to over 80\%.

\subsection{Parameter Sensitivity}
\label{sec:evaluation:paramChoices}

\satyam's groundtruth fusion and result evaluation algorithms have several parameters and \figref{fig:accuracy} presents results for the best parameter choice. We have analyzed the entire space of parameters to determine the parameters that \satyam performance most crucially depends upon in terms of accuracy, latency and cost.

Image classification is sensitive only to two parameters: $n_{min}$, the minimum number of results before aggregation can commence, and $\beta$, the fraction determining the super-majority. The upper graph in \figref{fig:param:image_classification} shows how classification accuracy varies as a function of these two parameters. Because the cost and latency of groundtruth collection varies with parameters, the lower graph shows the cost (blue bars) and the latency (red bars) for each parameter choice. From this, we can see that when $n_{min}\ge 3$  and $\beta \ge 0.7$ the accuracy does not improve significantly, however, cost and latency increase. This indicates that $n_{min}=3$ and $\beta = 0.7$ are good parameter choices, with high accuracy, while having moderate cost and latency. \looseness=-1

 
We have conducted similar analyses for video classification, counting, object detection (\figref{fig:param:localization}), segmentation, and tracking (\figref{fig:param:tracking}). Space constraints preclude a detailed discussion, but the key conclusions are:
(a) All job categories are sensitive to $n_{min}$, the minimum number of results before \satyam attempts to aggregate results;
(b) Each category is sensitive to one other parameter. For classification, this is the $\beta$ parameter that determines the super-majority criterion. For counting, it is the error tolerance $\epsilon$. For detection and tracking, it is $\eta_{cov}$, the fraction of corroborated groundtruth elements; and
(c) In each case, there exists a parameter settings at which provides good groundtruth performance at moderate cost and latency.

\section{Related Work}
\label{sec:related}

\parab{Image recognition using crowdtasking}
ImageNet training data for classification was generated using AMT, and uses majority voting for consensus~\cite{ImageNet}. Prior work~\cite{bbox_crowd12} has also shown crowdtasking to be successful for detection: unlike \sysname, in this work, quality control is achieved by using workers to rate other workers, and majority voting picks the best bounding box. These use one-off systems to automate HIT management and consensus, but do not consider payment management. \sysname achieves comparable performance to these systems but supports more vision tasks. Third party commercial crowdtasking systems exist to collect groundtruth for machine vision~\cite{crowdflower,spare5_website}. Other approaches have developed one-off systems built on top of AMT for more complex vision tasks, including feature generation for sub-class labeling~\cite{Deng13}, and sentence-level textual descriptions~\cite{VisualGenome,Datasift14}: more generally, future machine vision systems will need annotated groundtruth for other complex annotations including scene characterization, activity recognition, visual story-telling~\cite{CSsurvey} and we have left it to future work to extend \sysname to support these. \looseness=-1

\parab{Crowdtasking cost, quality, and latency} Prior work has extensively used multiple worker annotations and majority voting to improve quality~\cite{ImageNet,bbox_crowd12}. For binary classification tasks in a one-shot setting, lower cost solutions exist to achieve high quality~\cite{Karger11} or low latency~\cite{KrishnaError}. For top-$k$ classification (e.g., finding the $k$ least blurred images in a set) several algorithms can be used for improving crowdtasking consensus~\cite{Topk16}. Other work has explored this cost-quality tradeoff~\cite{Challenges16} in different crowd-tasking settings: de-aliasing entity descriptions~\cite{Waldo17,Khan16}, or determining answers to a set of questions~\cite{CrowdDQS17}. \sysname devises novel automated consensus algorithms for image recognition tasks based on the degree of pixel overlap between answers.

\parab{Crowdtasking platforms} Many marketplaces put workers in touch with requesters for freelance work~\cite{Guru,Freelancer,99design}, for coders~\cite{TopCoder}, for software testing~\cite{Mob4Hire,utest}, or for generic problem solving~\cite{Innocentive}. \satyam adds automation on top of an existing generic marketplace, AMT. Other systems add similar kinds of automation, but for different purposes. Turkit~\cite{turkit} and Medusa~\cite{Medusa} provide an imperative high-level programming language for human-in-the-loop computations and sensing respectively. Collaborative crowdsourcing~\cite{collabcrowd} automates the decomposition of more complex tasks into simpler ones, and manages their execution. 


\section{Conclusions}
\label{sec:concl}

In this paper, we have presented \satyam, a cloud-based platform for automating large-scale groundtruth collection for machine vision applications. \satyam's groundtruth matches that of existing ML benchmarks datasets, ML models retrained with \satyam are as good as those re-trained with benchmark datasets, and ML models fine-tuned with \satyam's groundtruth improve detection accuracy by up to 28\% in real deployments over pre-trained models. 

\newpage
\balance
\bibliography{references}
\bibliographystyle{sysml2019}

\end{document}